\documentclass{article}

\usepackage{amsmath, amssymb} 
\usepackage{enumerate}
\usepackage{pgf}
\usepackage{tikz}
\newcommand{\bc}{\begin{C}}
\newcommand{\ec}{\end{C}}
\newcommand{\be}{\begin{equation}}
\newcommand{\ee}{\end{equation}}

\newcommand{\claim}{\begin{Cl}}
\newcommand{\eclaim}{\end{Cl}}
\newcommand{\nb}{\begin{Nb}}
\newcommand{\nbe}{\end{Nb}}
\newcommand{\bl}{\begin{LE}}
\newcommand{\el}{\end{LE}}

\newtheorem{Cl}{Claim}

\newcommand{\bd}{\begin{Def}}
\newcommand{\ed}{\end{Def}}
\newcommand{\bt}{\begin{Th}}
\newcommand{\et}{\end{Th}}

\newtheorem{Th}{Theorem}
\newtheorem{LE}{Lemma}
\newtheorem{C}{Corollary}
\newtheorem{Nb}{Note}
\newtheorem{Def}{Definition}

\def\id{\rm id}
\def\ev{\rm ev}

\def\d{{\partial}}

\DeclareMathSymbol{\mlq}{\mathord}{operators}{``}
\DeclareMathSymbol{\mrq}{\mathord}{operators}{`'}

\begin{document}
 \title{Classical linear logic, cobordisms and categorical semantics of categorial grammars
}
\author{Sergey Slavnov
\\  National Research University Higher School of Economics
\\ sslavnov@yandex.ru\\} \maketitle

\begin{abstract}
We propose a categorial grammar based on classical multiplicative linear logic.

This can be seen as an extension of abstract categorial grammars (ACG) and is at least as expressive. However, constituents of {\it linear logic grammars (LLG)} are not abstract ${\lambda}$-terms, but simply tuples of words with labeled endpoints, we call them {\it multiwords}. At least, this gives a concrete and intuitive representation of ACG.

A key observation is that the class of multiwords has a fundamental algebraic structure. Namely, multiwords can be organized in a category, very similar to the category of topological cobordisms. This category is symmetric monoidal closed and compact closed and thus is a model of linear $\lambda$-calculus and classical linear logic. We think that this category is interesting on its own right.
In particular, it might provide categorical representation for other formalisms.

On the other hand, many models of language semantics are based on commutative logic or, more generally, on symmetric monoidal closed categories. But the category of {\it word cobordisms} is a category of language elements, which is itself symmetric monoidal closed and independent of any grammar. Thus, it   might  prove useful in understanding language semantics as well.
\end{abstract}

\section{Introduction}
A prototypical example of categorial grammar is Lambek grammars \cite{Lambek}. These are based on  logical {\it Lambek calculus}, which is, speaking in modern terms,  a noncommutative variant of (intuitionistic) linear logic \cite{Girard}. It is well known  that Lambek grammars  generate exactly the same class of languages as context-free grammars \cite{Pentus}.

However, it is agreed that context-free grammar are, in general, not sufficient for modeling natural language. Therefore linguists consider various more expressive formalisms. Lambek calculus is extended to different complicated  {\it multimodal}, {\it mixed commutative} and {\it mixed nonassociative} systems, see \cite{Moortgat}.
Many grammars operate with more complex constituents than just words. For example {\it displacement grammars} \cite{Morrill_Displacement},   extending Lambek grammars, operate on discontinuous tuples of words.

 Especially interesting (in the author's point of view) are {\it abstract categorial grammars (ACG)} \cite{deGroote}. Unlike Lambek grammars, these are based on a more intuitive and familiar {\it commutative} logic, namely,  the implicational fragment of  linear logic. Yet their expressive power is much stronger \cite{YoshinakaKanazawa}. This, however, comes with a certain drawback. The constituents are, basically, just linear $\lambda$-terms. It is not so easy to identify them with any elements of language. We should add  also that there exist {\it hybrid type logical grammars} \cite{KubotaLevine}, which extend ACG, mixing them with Lambek grammars.

 Finally, we  note, that, although the list of existing grammars seems sufficiently long, there exists a very interesting unifying approach of  \cite{Moot_comparing}. It turns out that many grammatical formalisms can be faithfully  represented as fragments of first order multiplicative intuitionistic linear logic {\bf MILL1}. This provides some common ground on which different systems can be compared. From the author's point of view it is quite remarkable that a unifying logic is, again, commutative.

 In this work we propose one more categorial grammar based on a commutative system, namely on classical linear logic. {\it Linear logic grammars (LLG)} of this paper can be seen as an extension of ACG to full multiplicative fragment. Although, as we just noted, the list of different formalisms is already sufficiently long, we think that our work deserves some interest at least for two reasons.

 First, unlike the case of ACG, constituents of LLG are very simple. They are tuples of words with labeled endpoints, we call them {\it multiwords}. Multiwords are directly identified as basic elements of language, and apparently they are somewhat easier to deal with than abstract $\lambda$-terms. ACG embed into LLG, so at least we give a concrete and intuitive {\it representation} of ACG. (We don't know if LLG have stronger expressive power as ACG, or just the same.)

 Second,
  we identify on the class of multiwords a fundamental algebraic structure. This structure is a {\it category} (in the mathematical, rather than linguistic sense of the word), which is {\it symmetric monoidal closed} and {\it compact closed}. It is this categorical structure that allows us representing linear $\lambda$-calculus and ACG, as well as classical linear logic. And, apparently, at least some other formalisms can be represented in this setting as well. Possibly,  this can give some common reference for different systems.

  We now discuss it in a greater detail.

\subsection{Algebraic considerations}
The algebraic structure underlying linguistic interpretations of Lambek calculus is that of a monoid.

Indeed, the set of words over a given alphabet is a free monoid under concatenation, and Lambek calculus can be interpreted as a logic of the poset of this monoid subsets (i.e. of formal languages). Typically, the sequent $$X_1,\ldots,X_n\vdash X$$ is interpreted as  subset inclusion: the concatenation of languages $X_1,\ldots,X_n$ is a sublanguage of $X$.

When constituents of a grammar are more complicated, such as word tuples, there is no unique concatenation, since tuples can be glued together in many ways. Thus the algebra is more complex.

We consider tuples of words with labeled endpoints, we call them {\it multiwords}. Multiwords can be conveniently represented as very simple directed graphs with labeled edges and vertices. They are glued together along matching labels on vertices.

For example, we have a multiword with two components
 $$
 
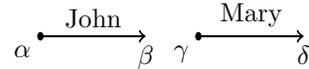
\tikz[xscale=.7]{

\draw[thick,->](0,0) -- (2,0);
\draw[thick,->](3,0) -- (5,0);
\draw [fill] (0,0) circle [radius=0.05];
\draw [fill] (3,0) circle [radius=0.05];
\node[above]at (1,0){John};
\node[above]at (4,0){Mary};
\node[below left] at (0,0){$\alpha$};
\node[below ]at (2,0){$\beta$};
\node[below left]at (3,0){$\gamma$};
\node[below ]at (5,0){$\delta$};
}
$$
and another multiword with one component.
$$
 
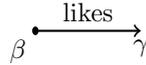
\tikz[xscale=.7]{
\draw[thick,->](0,0) -- (2,0);
\draw [fill] (0,0) circle [radius=0.05];
\node[above]at (1,0){likes};

\node[below left] at (0,0){$\beta$};
\node[below ]at (2,0){$\gamma$};
}
$$
These glue together and yield the following.
 $$
 
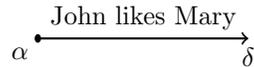
\tikz[xscale=.7]{

\draw[thick,->](0,0) -- (4,0);
\draw [fill] (0,0) circle [radius=0.05];
\node[above]at (2,0){John likes Mary};
\node[below left] at (0,0){$\alpha$};
\node[below ]at (4,0){$\delta$};

}
$$
The same multiword can be obtained by gluing a three-component multiword
$$
 
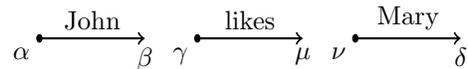
\tikz[xscale=.7]{

\draw[thick,->](0,0) -- (2,0);
\draw[thick,->](3,0) -- (5,0);
\draw[thick,->](6,0) -- (8,0);
\draw [fill] (0,0) circle [radius=0.05];
\draw [fill] (3,0) circle [radius=0.05];
\draw [fill] (6,0) circle [radius=0.05];
\node[above]at (1,0){John};
\node[above]at (4,0){likes};
\node[above]at (7,0){Mary};
\node[below left] at (0,0){$\alpha$};
\node[below ]at (2,0){$\beta$};
\node[below left]at (3,0){$\gamma$};
\node[below ]at (5,0){$\mu$};
\node[below left]at (6,0){$\nu$};
\node[below ]at (8,0){$\delta$};
}
$$
with another multiword
  
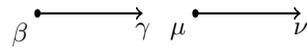
$$
 \tikz[xscale=.7]{
\draw[thick,->](0,0) -- (2,0);
\draw[thick,->](3,0) -- (5,0);
\draw [fill] (0,0) circle [radius=0.05];
\draw [fill] (3,0) circle [radius=0.05];
\node[below left] at (0,0){$\beta$};
\node[below ]at (2,0){$\gamma$};
\node[below left]at (3,0){$\mu$};
\node[below ]at (5,0){$\nu$};
}
$$
whose all components are empty.

Unfortunately, nothing precludes us from gluing words cyclically, and thus obtaining  cyclic sequences of letters with no endpoints. Consider gluing a word
$$
 
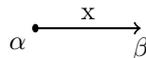
\tikz[xscale=.7]{

\draw[thick,->](0,0) -- (2,0);
\draw [fill] (0,0) circle [radius=0.05];
\node[above]at (1,0){x};
\node[below left] at (0,0){$\alpha$};
\node[below ]at (2,0){$\beta$};
}
$$
with a ``wrongly oriented'' one.
$$
 \tikz[xscale=.7]{
\draw[thick,->](0,-0.5) -- (2,-0.5);
\draw [fill] (0,-.5) circle [radius=0.05];
\node[above]at (1,-.5){y};

\node[below left] at (0,-0.5){$\beta$};
\node[below ]at (2,-0.5){$\alpha$};
}
$$

For consistency we have to allow also such {\it cyclic} or {\it singular} multiwords, which can be represented as closed loops.

Multiwords can be organized in a {\it monoidal category}, very similar to the category of {\it topological cobordisms} (see \cite{Baez}). Its objects, {\it boundaries}, are sets of vertex labels, and morphisms, {\it word cobordisms}, are (equivalence classes of) multiwords, composed by gluing.

Monoidal structure, ``tensor product'' is just disjoint union.

Thus, we shift from a non-commutative monoid  of words  to a symmetric monoidal category of word cobordisms.
(We find it amusing to abbreviate the latter term as {\it cowordism}.)

\subsection{Adding logic}
The category of cowordisms (over a given alphabet) is not only symmetric monoidal, but also {\it compact closed}, just as the category of cobordisms.
This makes it  a model of classical multiplicative  linear logic \cite{Seely}.

When interpreting logic in such a setting, logical consequence  does no longer correspond to subset inclusion.
A sequent $$X_1\ldots,X_n\vdash X$$ given {\it together with its derivation},  is now a particular cowordism  of type
$$X_1\otimes\ldots\otimes X_n\to X,$$
 which can be explicitly computed from the derivation.

Adding a {\it lexicon}, which is a finite set of non-logical axioms, i.e. cowordisms together with their typing specifications, we obtain a {\it linear logic grammar} (LLG).

Syntactic derivations from the lexicon directly translate to cowordisms, (which are just tuples of words).
This gives us a {\it linear logic grammar}; its language consists of all words that can be written as compositions of cowordisms in the lexicon and ``natural'' cowordisms coming from linear logic proofs.

Speaking more generally, with an LLG we get a subcategory of {\it cowordism types} generated by the grammar. This is, in general, no longer compact. It is, however, a categorical model of linear logic and linear $\lambda$-calculus.

Comparing with Lambek calculus, we shift    from a {\it poset} of formal languages to a {\it category} of cowordism types.

\subsection{Some wishful thinking on categorical semantics}
LLG are at least as expressive as abstract categorial grammars (on the string signature). Indeed, ACG are based on a conservative fragment of classical linear logic, so they  have direct translation to our setting. Thus, cowordisms and LLG provide a {\it concrete categorical model} of abstract categorial grammar.

In fact, cowordisms are essentially proof-nets, and passage from ACG to LLG is basically, a passage, from $\lambda$-terms to proof-nets.
Now,  forgetting about LLG, it seems reasonable that any  formalism admitting some version of proof-nets has a  representation in the category of cowordisms.  (It does not necessarily mean that such a representation is useful.) Possibly, this might provide some common, syntax-independent ground, i.e. a model, for different systems. This might be compared with representation of different systems in {\bf MILL1} in \cite{Moot_comparing}.

One of the main features making categorial grammars interesting is that they allow a bridge between language   syntax and  language semantics (see \cite{MootRetore}). Semantics is often modeled by means of a commutative logic, most notably, linear logic as in \cite{Dalrymple}. But the category of cowordisms itself is  a symmetric monoidal category of language elements, which is independent of any grammar. It  might prove  helpful for understanding this bridge.

An interesting approach is that of {\it categorical compositional distributional models of meaning (DisCoCat)}) \cite{CoeckeSadrzadehClark}, \cite{Coecke_Lmabek_vs_Lambek}. In DisCoCat it is proposed to model and analyze language semantics by a functorial mapping (``quantization'') of syntactic derivations in a categorial grammar to the (symmetric) compact closed category {\bf FDVec} of finite-dimensional vector spaces. The approach has been developed so far mainly  on the base of Lambek grammars or pregroup grammars (see \cite{Lambek_pregroups}), which are, from the category-theoretical point of view, {\it non-symmetric monoidal closed}. On the other hand,  the cowordism category is symmetric and compact closed, and in this sense it is a better mirror  of {\bf FDVec}. Thus it seems a more natural candidate for quantization. Possibly, cowordism representation may help to apply  ideas of DisCoCat to LLG or ACG, thus going beyond context-free languages.

\subsection{Structure of the paper}
The paper is reasonably self-contained. We assume, however, that the reader has some basic acquaintance with categories, in particular, with monoidal categories, see \cite{MacLane} for background.

In the first section we define the category of word cobordisms (cowordisms). In the second section we discuss  monoidal closed categories in general, and monoidal closed structures of cowordism categories in particular. Section 3 introduces linear logic, its categorical semantics and, finally, linear logic grammars. In Section 4, as an example, we show that multiple context-free grammars encode in LLG, and that every LLG with a {\it $\otimes$-free lexicon} generates a multiple context-free language. This result is similar to  the known result that all second order ACG generate multiple context-free languages \cite{Salvati}. The fifth section is the encoding of ACG to LLG. Finally, in the last section we show how LLG generates an NP-complete language. The purpose of this last piece is mainly  illustrative. We try to convince the reader that the geometric language of cowordisms is indeed intuitive and convenient for analysing language generation.

\section{Word cobordisms}
\subsection{Multiwords}
Let $T$ be a finite alphabet.
We denote the set of all finite words in $T$ as $T^*$.

For consistency of definitions we will also have to consider cyclic words.

We say that  two words in $T^*$ are {\it cyclically equivalent} if they differ by a cyclic permutation of letters.
A {\it cyclic word} over  $T$ is an equivalence class of cyclically equivalent words in $T^*$.

For $w\in T^*$  we denote the corresponding cyclic word  as $[w]$.

Observe that there exists a perfectly well-defined {\it empty cyclic word}.

\bd
A {\bf regular multiword} $M$ over an alphabet $T$ is a  finite directed  graph with  edges labelled by words in $T^*$, such that each vertex is adjacent
to exactly one edge (so that it is a perfect matching).
\ed

The {\it left}, respectively, {\it right boundary} of a multiword $M$ is the set of vertices of the underlying graph that are heads, respectively, tails of some edges.

We denote the left boundary of $M$ as $\d_lM$ and the right boundary, as $\d_rM$.

The {\it boundary} $\d M$ of $M$ is the set $\partial M=\d_lM\cup\d_rM$.

\bd
A {\bf multiword} $M$ over the alphabet $T$ is a pair $M=(M_0,M_c)$, where $M_0$, the {\bf regular} part, is a regular multiword over $T$, and $M_c$, the {\bf singular} or {\bf cyclic} part, is a finite multiset of cyclic words over $T$.
\ed

The boundaries $\d M$, $\d_l M$, $\d_r M$ of a multiword $M$ are defined as corresponding boundaries of its regular part $M_0$.

The multiword is {\it acyclic} or {\it regular} if its singular part is empty. Otherwise it is {\it singular}.

A multiword $M$ can be pictured geometrically as the edge-labelled graph $M_0$ and a bunch of isolated loops labelled by elements of $M_c$.
The underlying geometric object is no longer  a graph, but it is a topological space. It is even a manifold with boundary.
 In fact, we can equivalently define a multiword as a 1-dimensional compact oriented manifold with boundary (up to a boundary fixing  homeomorphism), whose connected components are labelled  by cyclic words, if they are closed, and by ordinary words otherwise.

\subsubsection{Gluing}
It should be clear from a geometric representation how to glue multiwords. We now give a boring accurate definition.

First, we define the {\it disjoint union} of multiwords in the most obvious way.

If $M=(M_0,M_c)$, $M'=(M_0',M_c')$ are multiwords then we define the  disjoint union $M\sqcup M'$ as the multiword
$$M\sqcup M'=(M_0\sqcup M_0',M_c\sqcup M_c').$$

Next we define {\it contraction}, which corresponds to elementary gluing.

Let $M$ be a multiword and $x\in\partial_lM$, $y\in \partial_rM$.

The {\it contraction} $M/\{x=y\}$  of $x$ and $y$ in $M$ is obtained by identifying $x$ with $y$ in the underlying graph and gluing the corresponding edges into one. The words labeling the edges are also glued, i.e. concatenated.

This means the following.

If vertices $x$, $y$ are not connected by an edge in $M_0$, then let $t$ be the tail of the unique edge adjacent to $x$ and $z$ be the head of the unique edge adjacent to $y$. Let $u$ be the word labeling $(x,t)$ and $v$ be the word labeling $(z,y)$. We construct a new edge-labelled graph $M_0'$ by removing $x$ and $y$ together with their adjacent edges from $M_0$ and drawing an edge from $(z,t)$. The new edge is labelled by the concatenation $vu$.

We put $M/\{x=y\}=(M_0',M_c)$.

If $x$ and $y$ are connected by an edge, let $w$ be its label. We remove $x$, $y$ and $(x,y)$ from $M_0$, which gives us the new edge-labelled graph
$M_0'$, and we add to $M_c$ the cyclic word $[w]$, which gives us the new multiset $M_c'$. We put $M/\{x=y\}=(M_0',M_c')$.
\smallskip

Note that iterated contractions commute.
\nb\label{associativity of contraction}
Let $M$ be a multiword, and $x_1,x_2\in \d_lM$, $y_1,y_2\in \d_rM$. Then
$$((M/\{x_1=y_1\})/\{x_2=y_2\}=((M/\{x_2=y_2\})/\{x_1=y_1\}.\quad \Box $$
\nbe
\smallskip

In view of the above  we can define {\it multiple contractions}.

\bd
Let $M$ be a multiword. Let
$$X\subseteq \d_lM,\quad Y\subseteq \d_rM,$$
and let
$\phi:X\to Y$ be a bijection.

 The {\bf contraction} $M/\{X\stackrel{}{\cong} Y\}$ of $X$ and $Y$ along $\phi$ in $M$ is defined by
 $$M/\{X\stackrel{}{\cong}Y\}=(\ldots(M/\{x_1=\phi(x_1)\})\ldots)/\{x_n=\phi(x_n)\},$$
 where $\{x_1,\ldots,x_n\}$ is any enumeration of elements of $X$.
\ed

 (We omit the bijection $\phi$ from notation, because it will be clear from the context.)

Now let two  multiwords $M$, $M'$ be given.

Assume that we have subsets
$$X_l\subseteq\d_l M,\quad X_r\subseteq\d_r M,$$
$$X_l'\subseteq\d_l M',\quad X_r'\subseteq\d_r M',$$
and two bijections
$$\phi:X_l\to X_r',\quad\phi':X_l'\to X_r.$$

Let $X, X'$ be the disjoint  unions $X=X_l\sqcup X_r$, $X'=X_l'\sqcup X_r'$.

The {\it gluing} $M\sqcup_{X\cong X'}M'$ of $M$ and $M'$ along $X$ and  $X'$ is defined as the multiple contraction
$$M\sqcup_{X,X'}M'=((M\sqcup M')/\{X_l\cong X_r'\})/\{X_l'\cong X_r\}.$$

\subsection{Category of word cobordisms}
\subsubsection{Cowordisms}
We  remarked above that multiwords can be represented geometrically as very simple manifolds with boundary. Manifolds with boundary give rise to the category of {\it cobordisms}, see \cite{Baez}. We are now going to define a similar category  of {\it word cobordisms}. We find it amusing to abbreviate the latter term as {\it cowordism}, and we will do so.

\bd A {\bf  boundary}  is a finite set $X$ equipped with a partition $X=X_l\cup X_r$ into two disjoint subsets.
\ed

Now,  we want to look at a multiword $M$ as a morphism between boundaries. For that, we need to understand which part of $\d M$ is the input, and which is the output. This leads to the following definition.

\bd
Let $X$, $Y$ be boundaries.

A {\bf cowordism} $$\sigma:X\to Y$$
 over an alphabet $T$ from $X$ to $Y$ is  a triple
 $$\sigma=(M,\phi_l,\phi_r),$$
 where $M$ is a multiword over $T$ together with
two bijective {\bf labeling functions} $$\phi_l:Y_l\sqcup X_r\to\d_lM,\quad\phi_r:Y_r\sqcup X_l \to\d_rM.$$
\ed

A cowordism is {\it regular} if its underlying multiword is regular. Otherwise the cowordism is {\it singular}.

 For our purposed it is necessary to identify  cowordisms that differ by inessential relabeling of boundaries. Therefor we supply our definition of a cowordism  with a  definition of {\it cowordism equality}.

\bd
Two cowordisms $\sigma=(M,\phi_l,\phi_r)$ and $\sigma'=(M',\phi_l',\phi_r')$ are  {\bf equal}, if their singular part coincide,
$$M_c=M_c',$$
 and there is a pair of bijections
$$\psi_l:\d_lM\to\d_lM',\mbox{ }\psi_r:\d_rM\to\d_rM'$$ inducing an edge-labeled graph isomorphism of the regular parts,  such  that
$$\phi_l'=\psi_l\circ\phi_l,\mbox{ }\phi_r'=\psi_r\circ\phi_r.$$
\ed

In the sequel we will systematically abuse notation and denote a cowordism and its underlying multiword with the same letter.

Note, however, that,  generally speaking, a cowordism and a multiword are two different structures. In particular, we can have two different non-equal multiwords representing the same cowordism (see the definition of cowordism equality above).

%
%
%
%
%
%
%

We are going to organise cowordisms into a compact closed category (to be discussed below).
Since cowordisms, by definition, have geometric representation, it is natural to adapt the {\it pictorial language} (see \cite{Selinger}) used for such categories.

We can  depict an abstract cowordism $\sigma:X\to Y$ schematically as a box with incoming and outgoing wires, like the following.

 $$
 
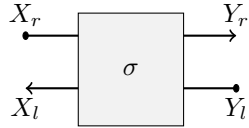
\tikz[xscale=.7]{
        \draw[draw=black,fill=gray!10](0,0)rectangle(2,1.5);\node at(1,.75){$\sigma$};
 \draw[thick,-](2,.5)--(3,.5);
 \node[above] at(3,1.2) {$Y_r$};
 \draw[thick,->](2,1.2)--(3,1.2);
 \node[below] at(3,.5) {$Y_l$};

 \node[below ] at(-1,.5) {$X_l$};
 \node[above ] at(-1,1.2) {$X_r$};
 \draw [fill] (-1,1.2) circle [radius=0.05];
 \draw [fill] (3,.5) circle [radius=0.05];
 \draw[thick](-1,1.2)--(0,1.2);
 \draw[thick,<-](-1,.5)--(0,.5);
 }.$$

Or, using fewer labels on the wires, like the following.

  $$
 
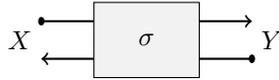
\tikz[xscale=.7]{
        \draw[draw=black,fill=gray!10](0,0)rectangle(2,1);\node at(1,.5){$\sigma$};
 \draw[thick,-](2,.25)--(3,.25);

 \draw[thick,->](2,.75)--(3,.75);
 \node [right] at(3,.5) {$Y$};

 \node[left ] at(-1,.5) {$X$};

 \draw [fill] (-1,.75) circle [radius=0.05];
 \draw [fill] (3,.25) circle [radius=0.05];
 \draw[thick](-1,.75)--(0,.75);
 \draw[thick,<-](-1,.25)--(0,.25);
 }.$$

(Of course for a concrete $\sigma$ there are as many wires as there are points in the boundaries $X$, $Y$.)

\subsubsection{Composition}
Cowordisms are composed simply by gluing multiwords along matching boundary parts.

In the pictorial language of boxes and wires, given two cowordisms
$$\sigma:X\to Y,\quad\tau:Y\to Z,$$
the composition $\tau\circ\sigma$ is represented in a most natural way.

 $$
 
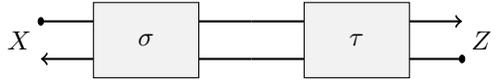
\tikz[xscale=.7]{
        \draw[draw=black,fill=gray!10](0,0)rectangle(2,1);\node at(1,.5){$\sigma$};
 \draw[thick,-](2,.25)--(3,.25);
 \draw[thick,-](2,.75)--(3,.75);

 \node[left ] at(-1,.5) {$X$};
  \draw [fill] (-1,.75) circle [radius=0.05];
 \draw[thick](-1,.75)--(0,.75);
 \draw[thick,<-](-1,.25)--(0,.25);

  \draw[draw=black,fill=gray!10](4,0)rectangle++(2,1);\node at(5,.5){$\tau$};
 \draw[thick,-](6,.25)--(7,.25);

 \draw[thick,->](6,.75)--(7,.75);
 \node[right] at(7,.5) {$Z$};

 \draw [fill] (7,.25) circle [radius=0.05];
 \draw[thick](3,.75)--(4,.75);
 \draw[thick,-](3,.25)--(4,.25);
 }.$$

An accurate definition is as follows.

Let $X$, $Y$, $Z$ be boundaries, and
$$\sigma=(\sigma,\phi_l,\phi_r),\quad \tau=(\tau,\psi_l,\psi_r)$$ be  cowordisms from $X$ to $Y$ and from $Y$ to $Z$ respectively.

Let $\rho=\sigma\sqcup\tau$.

We have the injective maps
$$\xi_l:Y_l\sqcup Y_r\to \d_l\rho,\mbox{ }\xi_r: Y_r\sqcup Y_l\to \d_r\rho$$
obtained from restrictions of $\phi_l\sqcup\psi_l$, $\psi_r\sqcup\phi_r$ respectively.

Denote the image of $\xi_l$ as $I_l$ and  the image of $\xi_r$ as  $I_r$.

The {\it composition} $\tau\circ\sigma$ is defined as the gluing of $\tau$ and $\sigma$ along $I_l$ identified with $I_r$ by means of bijection $\xi_r^{-1}\circ\xi_l:I_l\cong I_r$, i.e.
$$\tau\circ\sigma=
(\sigma\sqcup\tau)/\{I_l\stackrel{}{\cong}I_r\}.$$

Restrictions of $\psi_l\sqcup\phi_l$ to $Z_l\sqcup X_r$ and of $\psi_r\sqcup\phi_r$ to $Z_r\sqcup X_l$ provide necessary bijections
$$Z_l\sqcup X_r\cong \d_l(\tau\circ\sigma),\mbox{ }Z_r\sqcup X_l\cong\d_r(\tau\circ\sigma),$$
which makes the constructed multiword a cowordism from $X$ to $Z$.

It follows from Note \ref{associativity of contraction} and definition of cowordism equality that composition is associative.

\subsubsection{Identities}
In order to construct a category we only need to find identities.

Let $X$ be a boundary.

The {\it identity cowordism} $\id_X$ is constructed as follows.

Take two copies of $X$ and then draw a directed edge from each point of $X_r$ in the first copy  to its image in the second copy and from each point of $X_l$ in the second copy to its image in the first copy. Label every constructed edge with the empty word. This gives us an acyclic multiword with the left and right boundaries  isomorphic to $X_r\sqcup X_l$.

In the pictorial language, $\id_X$ looks as follows.
  $$
 \tikz[xscale=.7]{

\draw[thick,->](0,0) -- (2,0);
\draw[thick,<-](0,-0.5) -- (2,-0.5);

\node[left] at (0,-0.25) {$X$};
\node[right] at (2,-0.25) {$X$};

 \draw [fill] (0,0) circle [radius=0.05];
  \draw [fill] (2,-0.5) circle [radius=0.05];

 }$$

\smallskip

It is immediate now that the following is well defined.

\bd
The category ${\bf Cow}_T$ of cowordisms over the alphabet $T$
has boundaries as objects and cowordisms over $T$ as morphisms.
\ed

\subsection{Over the empty alphabet}
Note that even when the alphabet  is empty, the category of cowordisms is nontrivial.
In fact, it becomes literally the category of oriented 1-dimensional cobordisms.

In the sequel we will use  the term {\it cobordism} for a cowordism over the empty alphabet, and
 denote $${\bf Cow_\emptyset}={\bf Cob}.$$

Given two boundaries $X,Y$ and a cowordism $\sigma:X\to Y$ over some alphabet $T$, we define the {\it pattern} of $\sigma$ as the cobordism from $X$ to $Y$ obtained by erasing from $\sigma$ all letters.

\section{Cowordisms and monoidal closed categories}
\subsection{Structure of cowordisms category}
The category of cowordisms has a rich structure (which it inherits, in fact, from the underlying category of cobordisms).

It is a {\it symmetric monoidal closed, $*$-autonomous}, and {\it compact closed category}, which makes it a model of linear $\lambda$-calculus and of classical multiplicative linear logic.

\subsubsection{Monoidal structure}
First, the operation of disjoint union makes this category {\it monoidal}.

The {\it tensor product} $\otimes$ on ${\bf Cow}_T$ is defined both on objects and morphisms as the disjoint union.

The {\it monoidal unit} ${\bf 1}$ is the empty boundary,
$${\bf 1}={\bf 1}_r={\bf 1}_l=\emptyset.$$

Obviously, tensor product of cowordisms is associative up to a natural transformation.

In order to avoid very cumbersome notations we will, as is quite customary in literature, treat the category of cowordisms as {\it strict monoidal}. That is we will write $X\otimes Y \otimes Z$ without brackets, as if the associativity isomorphisms were strict equalities. Similarly, we will usually identify ${\bf 1}\otimes X$ and $X\otimes {\bf 1}$ with $X$.
This is legitimate, because any monoidal category is equivalent to a strict monoidal category, see \cite{MacLane}, Chapter VII for details.

In the pictorial language, given two cowordisms
$$\sigma:X\to Y,\quad\tau:Z\to T,$$
we depict the tensor product $\sigma\otimes \tau$ as two disjoint boxes.

$$
 
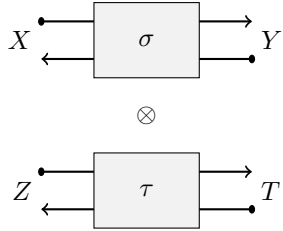
\tikz[xscale=.7]{
         \draw[draw=black,fill=gray!10](0,0)rectangle(2,1);\node at(1,.5){$\sigma$};
 \draw[thick,-](2,.25)--(3,.25);
 \node[right] at(3,.5) {$Y$};
 \draw[thick,->](2,.75)--(3,.75);

 \node[left ] at(-1,.5) {$X$};

 \draw [fill] (-1,.75) circle [radius=0.05];
 \draw [fill] (3,.25) circle [radius=0.05];
 \draw[thick](-1,.75)--(0,.75);
 \draw[thick,<-](-1,.25)--(0,.25);

 \node at(1,-.5){$\otimes$};

         \draw[draw=black,fill=gray!10](0,-2.)rectangle++(2,1.);\node at(1,-1.5){$\tau$};
 \draw[thick,->](2,-1.25)--(3,-1.25);
 \node[right] at(3,-1.5) {$T$};
 \draw[thick,-](2,-1.75)--(3,-1.75);

 \node[left ] at(-1,-1.5) {$Z$};

 \draw [fill] (-1,-1.25) circle [radius=0.05];
 \draw [fill] (3,-1.75) circle [radius=0.05];
 \draw[thick,<-](-1,-1.75)--(0,-1.75);
 \draw[thick](-1,-1.25)--(0,-1.25);
 }.$$

For an abstract cowordism $\sigma$ of the form
$$\sigma:X_1\otimes\ldots\otimes X_n\to Y_1\otimes\ldots\otimes Y_m, $$
it is convenient to depict $\sigma$ as a box with different slots for different tensor factors, as follows.
$$
 
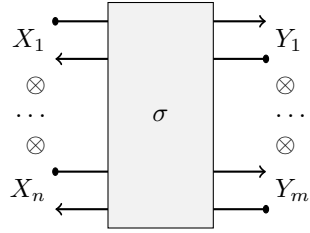
\tikz[xscale=.7]{
         \draw[draw=black,fill=gray!10](0,-2)rectangle++(2,3);\node at(1,-.5){$\sigma$};
 \draw[thick,-](2,.25)--(3,.25);
 \node[right] at(3,.5) {$Y_1$};
 \draw[thick,->](2,.75)--(3,.75);

 \node[left ] at(-1,.5) {$X_1$};
 \node[left ] at(-1,-.1) {$\otimes$};
\node[right ] at(3,-.1) {$\otimes$};
 \draw [fill] (-1,.75) circle [radius=0.05];
 \draw [fill] (3,.25) circle [radius=0.05];
 \draw[thick](-1,.75)--(0,.75);
 \draw[thick,<-](-1,.25)--(0,.25);

 \node [left] at(-1,-.5){$\ldots$};
\node [right] at(3,-.5){$\ldots$};

 \draw[thick,->](2,-1.25)--(3,-1.25);
 \node[right] at(3,-1.5) {$Y_m$};
 \draw[thick,-](2,-1.75)--(3,-1.75);

\node[left ] at(-1,-.9) {$\otimes$};
\node[right ] at(3,-.9) {$\otimes$};
 \node[left ] at(-1,-1.5) {$X_n$};

 \draw [fill] (-1,-1.25) circle [radius=0.05];
 \draw [fill] (3,-1.75) circle [radius=0.05];
 \draw[thick,<-](-1,-1.75)--(0,-1.75);
 \draw[thick](-1,-1.25)--(0,-1.25);
 }.$$

When the cowordism $\sigma$ is of the form
$$\sigma:{\bf 1}\to X_1\otimes\ldots\otimes X_n$$
It is natural to represent it without wires on the left as follows.

$$
 
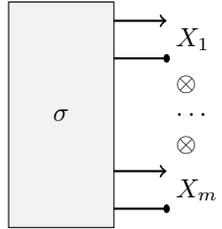
\tikz[xscale=.7]{
         \draw[draw=black,fill=gray!10](0,-2)rectangle++(2,3);\node at(1,-.5){$\sigma$};
 \draw[thick,-](2,.25)--(3,.25);
 \node[right] at(3,.5) {$X_1$};
 \draw[thick,->](2,.75)--(3,.75);

\node[right ] at(3,-.1) {$\otimes$};

 \draw [fill] (3,.25) circle [radius=0.05];

\node [right] at(3,-.5){$\ldots$};

 \draw[thick,->](2,-1.25)--(3,-1.25);
 \node[right] at(3,-1.5) {$X_m$};
 \draw[thick,-](2,-1.75)--(3,-1.75);

\node[right ] at(3,-.9) {$\otimes$};

 \draw [fill] (3,-1.75) circle [radius=0.05];
 }.$$

\subsubsection{Symmetry}
The above monoidal structure  is also symmetric.

The {\it symmetry transformation}
$$s_{X,Y}:X\otimes Y\to Y\otimes X$$
is given for any boundaries $X$, $Y$ by the following cowordism.

Take a copy of $X\sqcup Y$ and a copy of $Y\sqcup X$. For each $x\in X_r$ draw a directed edge from the image of $x$ in  $X\sqcup Y$ to the image of $x$ in $Y\sqcup X$, similarly for each $y\in Y_r$. Then for each
$x\in X_l$ draw a directed edge from the image of $x$ in  $Y\sqcup X$ to the image of $x$ in $X\sqcup Y$, similarly for each $y\in Y_l$.
Label each constructed edge with the empty word. This gives an acyclic multiword, which is a cowordism from $X\otimes Y$ to $Y\otimes X$ in the obvious way.

In the pictorial language symmetry is the following.

  $$
 
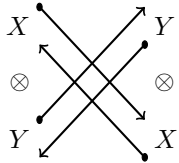
\tikz[xscale=.7]{

\draw[thick,->](0,0) -- (2,-1.5);
\draw[thick,<-](0,-.5) -- (2,-2);
\draw[thick,->](0,-1.5) -- (2,0);
\draw[thick,<-](0,-2.) -- (2,-.5);

\node[right] at (2,-1) {$\otimes$};
\node[left] at (0,-1) {$\otimes$};

\node[right] at (2,-.25) {$Y$};
\node[right] at (2,-1.75) {$X$};
\node[left] at (0,-.25) {$X$};
\node[left] at (0,-1.75) {$Y$};
 \draw [fill] (0,0) circle [radius=0.05];
\draw [fill] (2,-.5) circle [radius=0.05];
  \draw [fill] (0,-1.5) circle [radius=0.05];
\draw [fill] (2,-2.) circle [radius=0.05];
 }$$

\nb
The above defined tensor product, monoidal unit and symmetry make ${\bf Cow}_T$ a symmetric monoidal category. $\Box$
\nbe

\subsubsection{Duality and internal homs}
The category of cowordisms also has a well-behaved {\it contravariant duality} $(.)^\bot$, defined
 by switching  left  and right.

Let $X=X_r\cup X_l$ be a boundary.

The {\it dual} $X^\bot$ of $X$  is defined by
$$X^\bot=X,\quad (X^\bot)_r=X_l,\quad(X^\bot)_l=X_r.$$

On morphisms, duality amounts to relabeling boundary points.

Let $\sigma:X\to Y$ be a cowordism.

By definition $\sigma$ is a multiword $\sigma$ together with two  labeling functions $$\phi_l:Y_l\sqcup X_r\to\d_l\sigma,\quad\phi_r:Y_r\sqcup X_l \to\d_r\sigma.$$
Let $$s_{r,l}:X_r\sqcup Y_l\to Y_l\sqcup X_r,\quad s_{l,r}:X_l\sqcup Y_r\to Y_r\sqcup X_l$$
be the natural bijections.

Then the triple $$\sigma^\bot=(\sigma,\phi_r\circ s_{l,r},\phi_l\circ s_{r,l})$$
is a cowordism from $Y^\bot$ to $X^\bot$.

In the pictorial language, given a cowordism $\sigma:X\to Y$, the dual cowordism $\sigma^\bot$ looks as follows.

 $$
 
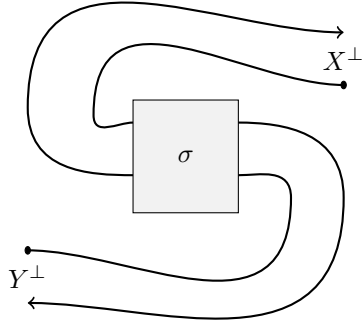
\tikz[xscale=.7]{
        \draw[draw=black,fill=gray!10](0,0)rectangle(2,1.5);\node at(1,.75){$\sigma$};

 \draw [fill] (4,1.7) circle [radius=0.05];
 \draw [fill] (-2,-.5) circle [radius=0.05];

 \draw[thick,-](2,.5) to  [out=0,in=90] (3,.2) to  [out=-90,in=0] (-2,-.5);

\draw[thick,->](2,1.2) to  [out=0,in=90] (4,.2) to  [out=-90,in=0] (-2,-1.2);

\draw[thick,-](4,1.7) to  [out=180,in=90] (-.75,1.3) to  [out=-90,in=180] (0,1.2);

\draw[thick,<-](4,2.4) to  [out=180,in=90] (-2,1.4) to  [out=-90,in=-180] (0,.5);
\node []at (4,2.05) {$X^\bot$};
\node []at (-2,-.85) {$Y^\bot$};

 }.$$

\nb
The above defined duality is a contravariant functor commuting with the tensor product: $(X\otimes Y)^\bot\cong X^\bot\otimes Y^\bot$. $\Box$
\nbe
\smallskip

Tensor and duality equip ${\bf Cow}_T$  with a very rich categorical structure that we discuss in the next section.
%

\subsection{Zoo of monoidal closed categories}
\bd {\bf Monoidal closed category} ${\bf C}$ is a symmetric monoidal category ${\bf C}$  equipped with a bifunctor $\multimap$, contravariant in the first entry and covariant in the second entry,
such that there exists a natural bijection
\be\label{monoidal closure}
Hom(X\otimes Y, Z)\cong Hom(X,Y\multimap Z).
\ee
\ed

The functor $\multimap$ in the above definition is called {\it internal homs functor}.

\bd
 \cite{Barr} {\bf $*$-Autonomous category} ${\bf C}$ is a symmetric monoidal category ${\bf C}$  equipped with a
contravariant functor $(.)^\bot$, such that there is a natural isomorphism
$$A^{\bot\bot}\cong A$$
and a natural bijection
$$
Hom(X\otimes Y, Z)\cong Hom(X,(Y\otimes Z^\bot)^\bot.
$$
\ed

Duality $(.)^\bot$ equips a $*$-autonomous category with a second monoidal structure. The {\it cotensor product} $\wp$ is defined by
$$X\wp Y=(X^\bot\otimes ^\bot)^\bot.$$
The neutral object for the cotensor product is
$$\bot={\bf 1}^\bot.$$

Any $*$-autonomous category is monoidal closed. The internal homs functor is defined by
$$X\multimap Y=X^\bot\wp Y.$$

Note that we have a natural isomorphism
\be\label{duality as implication}
X^\bot\cong X\multimap\bot.
\ee

\bd\cite{KellyLaplaza}
A {\bf compact closed} or, simply, {\bf compact category}  is a $*$-autonomous category for which duality commutes with tensor, i.e. such that
$$X\wp Y\cong X\otimes Y,\quad {\bf 1}\cong \bot.$$
\ed

For compact categories it is convenient to define  internal homs by\be\label{internal homs in compacts}
X\multimap Y=X^\bot\otimes Y,
\ee

 A prototypical example of a compact category is the category of finite-dimensional vector spaces with the usual tensor product and algebraic duality. Note, however, that in this case, and, in general, in the algebraic setting, duality is denoted as a star $(.)^*$.
 Another example of a compact category widely used in mathematics and important for our discussion is the category of  cobordisms.

\nb
The category of cowordisms is compact closed (hence monoidal closed and $*$-autonomous).
\nbe
{\bf Proof} exercise. $\Box$
\smallskip

Compact structure provides a lot of important maps and constructions. A short and readable introduction into the subject  can be found, for example, in \cite{AbramskyCoecke}.

We pick some necessary bits in the next section.

\subsubsection{Names}
     Let ${\bf C}$ be a monoidal closed category.

For any morphism $$\sigma:A\to B$$
correspondence (\ref{monoidal closure}) together with the isomorphism
$$A\cong{\bf 1}\otimes A$$ yields the morphism
$$\ulcorner\sigma\urcorner:{\bf 1}\to A\multimap B,$$
sometimes called the {\it name}
of $\sigma$.

In the case of cowordisms, the name $\ulcorner\sigma\urcorner:{\bf 1}\to A\multimap B\cong A^\bot\otimes B$ of a cowordism $\sigma:A\multimap B$ can be depicted as follows.
$$
 
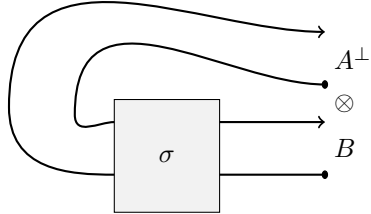
\tikz[xscale=.7]{
        \draw[draw=black,fill=gray!10](0,0)rectangle(2,1.5);\node at(1,.75){$\sigma$};

 \draw [fill] (4,1.7) circle [radius=0.05];
 \draw [fill] (4,.5) circle [radius=0.05];

 \draw[thick,-](2,.5) --(4,.5);

\draw[thick,->](2,1.2)--(4,1.2);

\draw[thick,-](4,1.7) to  [out=180,in=90] (-.75,1.3) to  [out=-90,in=180] (0,1.2);

\draw[thick,<-](4,2.4) to  [out=180,in=90] (-2,1.4) to  [out=-90,in=-180] (0,.5);
\node [right]at (4,2.05) {$A^\bot$};
\node [right]at (4,.85) {$B$};
\node [right]at (4,1.45) {$\otimes$};
 }.$$

\subsubsection{Applications}
As before, let ${\bf C}$ be a monoidal closed category.

For any two objects $A,B$, correspondence (\ref{monoidal closure}) composed with symmetry applied to $\id_{A\multimap B}$ yields the {\it evaluation} morphism
$$\ev_{A,B}:A\otimes(A\multimap B)\to B.$$

In a compact closed case, where we have identifications (\ref{internal homs in compacts}), evaluation is especially simple.

We have the natural {\it pairing map} $$\epsilon_{A}:A\otimes A^\bot\to{\bf 1},$$ usually called {\it counit}, and evaluation can be computed as $$\ev_{A,B}=\epsilon_A\otimes\id_B.$$

In the case of cowordisms the pairing $\epsilon_A$ has the following shape (remember that $A^\bot_r=A_l$ and $A^\bot_l=A_r$).

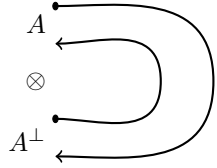
$$
 \tikz[xscale=.7]{

 \draw[thick,->](0,0) to  [out=0,in=90] (3,-1.) to  [out=-90,in=0] (0,-2);
\node[below left ] at (0,0){$A$};
\draw [fill] (0,0) circle [radius=0.05];

\draw [fill] (0,-1.5) circle [radius=0.05];
\draw[thick,<-](0,-0.5) to  [out=0,in=90] (2,-1) to  [out=-90,in=0] (0,-1.5);

\node [below left]at (0,-1.5) {$A^\bot$};
\node[left] at(0,-1) {$\otimes$};
 }$$

The evaluation $\ev_{A,B}$, accordingly, is pictured as follows.

 $$
 
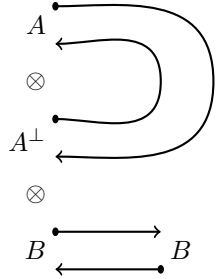
\tikz[xscale=.7]{

 \draw[thick,->](0,0) to  [out=0,in=90] (3,-1.) to  [out=-90,in=0] (0,-2);
\node[below left ] at (0,0){$A$};
\draw [fill] (0,0) circle [radius=0.05];

\draw [fill] (0,-1.5) circle [radius=0.05];
\draw[thick,<-](0,-0.5) to  [out=0,in=90] (2,-1) to  [out=-90,in=0] (0,-1.5);
\node [below left]at (0,-1.5) {$A^\bot$};

\node[left] at(0,-1) {$\otimes$};
\node[left] at(0,-2.5) {$\otimes$};

\draw[thick,->] (0,-3)--(2,-3);
\draw[thick,<-] (0,-3.5)--(2,-3.5);
\node [below left]at (0,-3) {$B$};
\node [below right]at (2,-3) {$B$};
\draw [fill] (0,-3) circle [radius=0.05];
\draw [fill] (2,-3.5) circle [radius=0.05];
 }$$

Now given two morphisms $$\tau:{\bf 1}\to A,\quad\sigma:{\bf 1}\to A\multimap B,$$
we can define the {\it application}
$$(\sigma\cdot\tau)_A:{\bf 1}\to B$$
of $\sigma$ to $\tau$ as
$$(\sigma\cdot\tau)_A=\ev_{A,B}\circ(\tau\otimes\sigma).$$

The following property holds for any monoidal closed category.
\nb\label{application}
For any two morphisms
$$\tau:{\bf 1}\to A,\quad \sigma: A\to B,$$
 it holds that
$$\ulcorner\sigma\urcorner\cdot\tau=\sigma\circ\tau.\mbox{ } \Box$$
\nbe
\smallskip

In the case of cowordisms, the  property is evident from geometric representation.

\subsubsection{Partial pairing}
Now let ${\bf C}$ be a $*$-autonomous category.

For any objects $A,B, C,D$ there is a natural {\it linear distributivity} morphism \cite{CocketSeely}
\be\label{linear distributivity}
\delta_{A,B,C,D}:(A\wp B)\otimes (C\wp D)\to A\wp (B\otimes C)\wp D.
\ee

In a compact closed case, where cotensor and tensor can be identified, linear distributivity is just associativity of tensor product.

Using linear distributivity, for any two morphisms
 $$\tau:{\bf 1}\to A\wp U,\quad\sigma:{\bf 1}\to U^\bot\wp B,$$
we can define the {\it  partial pairing}
$$\langle\tau,\sigma\rangle_U:{\bf 1}\to A\wp B$$
 of $\tau$ and $\sigma$ over $U$
 by
$$
\langle\tau,\sigma\rangle_U=(\id_A\wp \epsilon_{U}\wp\id_B)\circ\delta_{A,U,U^\bot,B}\circ(\tau\otimes\sigma).
$$

In the case of cowordisms, given two cowordisms
$$\sigma:{\bf 1}\to A\otimes U,\quad\tau:{\bf 1}\to U^\bot\otimes B,$$
the partial pairing $\langle\tau,\sigma\rangle_U$ has the following shape.

$$
 
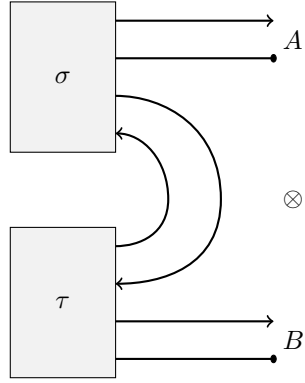
\tikz[xscale=.7]{
         \draw[draw=black,fill=gray!10](0,0)rectangle++(2,2);\node at(1,1){$\sigma$};
 \draw[thick,-](2,1.25)--(5,1.25);
 \node[right] at(5,1.5) {$A$};
 \draw[thick,->](2,1.75)--(5,1.75);

 \draw[thick,->](2,.75) to  [out=0,in=90] (4,-.625) to  [out=-90,in=0] (2,-1.75);

\draw[thick,<-](2,.25) to  [out=0,in=90] (3,-.625) to  [out=-90,in=0] (2,-1.25);

 \draw [fill] (5,1.25) circle [radius=0.05];

         \draw[draw=black,fill=gray!10](0,-3.)rectangle++(2,2);\node at(1,-2){$\tau$};
\draw[thick,->](2,-2.25)--(5,-2.25);
  \draw[thick,-](2,-2.75)--(5,-2.75);
 \node[right] at(5,-2.5) {$B$};
\node[right] at(5,-.625) {$\otimes$};

 \draw [fill] (5,-2.75) circle [radius=0.05];

 }.$$

Partial pairing can be understood as a symmetrized  composition, as the following observation shows.

\nb\label{partial pairing}

 For all morphisms
  $$\tau:A\to B,\quad\sigma:B\to C$$ it holds that
    $$\ulcorner\sigma\circ\tau\urcorner=\langle\ulcorner\tau\urcorner,\ulcorner\sigma\urcorner\rangle_{B}. \quad\Box$$
\nbe

\subsection{Categories of cowordism types}
We know discuss subcategories of ${\bf Cow}_T$, which are no longer compact, but are monoidal closed. They will be helpful for understanding categorial grammars considered in this paper.

\bd
Given a boundary $X$,  a {\bf cowordism type} over an alphabet $T$ or, simply, a {\bf type} on the boundary $X$ is a set of cowordisms over $T$ from ${\bf 1}$ to $X$.

A set of cowordisms over the alphabet $T$ is a {\bf cowordism type} or, simply, a {\bf type}, if it is a type on some boundary.
\ed

Given a type $A$, we denote the corresponding boundary as $\d A$.


\bd Given two cowordism types $A,B$ over the same alphabet,  a cowordism $$\sigma:\d A\to \d B$$ is a {\bf morphism of types} $$\sigma:A\to B$$ if for any
$\tau\in A$ it holds that $\sigma\circ \tau\in B$.
\ed


Obviously, morphisms of types compose, and identity cowordisms are morphisms of types. So, types over an alphabet $T$ form a category. We denote it as  ${\bf Types}_T$.


Categories of types inherit symmetrical monoidal, and even monoidal closed structure of ${\bf Cow}_T$.

For two types $A,B$ we define the {\it tensor product type} $A\otimes B$ as the type on the  tensor product of boundaries,
$$\d(A\otimes B)=\d A \otimes \d B,$$ given by
$$A\otimes B=\{\sigma\otimes \tau|\mbox{ }\sigma\in A,\tau\in B\}.$$


We define the {\it internal homs type} $A\multimap B$ as the type on the boundary
$$\d(A\multimap B)=\d A\multimap\d B=(\d A)^\bot\otimes\d B$$
given by
$$A\multimap B=\{\sigma|\mbox{ }\forall\tau\in A\mbox{ }\sigma\cdot\tau\in B\}.$$

Elements of  $A\multimap B$ are precisely all names of cowordisms which are morphisms of types $A$ and $B$.


The {\it unit type} ${\bf 1}$ is the type on the empty boundary that contains only the empty cowordism $\emptyset$.

\nb
The category ${\bf Types}_T$ of cowordism types  is symmetric monoidal closed.

 The forgetful functor $${\bf Types}_T\to {\bf Cow}_T$$
  which send each type $A$ to the boundary $\d A$ and is identity on morphisms preserves monoidal closed structure. $\Box$
\nbe

\subsubsection{Cowordisms of a formal language}\label{cowordism of lang}
Let $L$ be a formal language in the alphabet $T$.
Without loss of generality we assume that the symbol $\star$ is not in $T$. Let
$$T'=T\cup\{\star\}.$$

We define  the type $\bot$ over $T'$ on the empty boundary as the set of cyclic words
$$\bot=\{[w\star]\mbox{ }|w\in L\},$$
where each cyclic word is seen as a singular cowordism.

Now for any type $A$ over $T'$ we define the {\it dual} $A^\bot$ of  $A$ (with respect to $L$) as the type
$$A^\bot=A\multimap\bot.$$

We say that the type $A$ is a {\it closed type (of the language $L$)} if $A=A^{\bot\bot}$ (using the identification $(\d A)^{\bot\bot}\cong\d A$ on the level of boundaries).

 Closed types of $L$ form a (full) subcategory of ${\bf Types}_{T'}$, which we denote as ${\bf CTypes}_L$.

  The category ${\bf CTypes}_L$  is, in fact,  $*$-autonomous.

It is easy to see that for all closed types $A$, $B$, the type $A\multimap B$ is closed. Also the types
$\bot$, ${\bf 1}$ are closed with
$${\bf 1}=\bot^\bot.$$
In general, we have the following.

\nb
A type $A$ is closed iff $A\cong B^\bot$ for some type $B$ on the boundary $(\d A)^\bot$.

There is a contravariant functor
$$(.)^\bot:{\bf Types}_T\to {\bf CTypes}_L$$
sending a type $A$ to the type $A^\bot$ and a cowordism $\sigma$, to the cowordism $\sigma^\bot$.
$\Box$
\nbe
\smallskip

In particular, if $A$ is a type, then we can complete it to the  type $Cl(A)$ on the same boundary $\d A$, defined as
$$Cl(A)=A^{\bot\bot}$$ with the usual identification
\be\label{bidal identification}
\d A\cong(\d A)^{\bot\bot}.
\ee
We say that $Cl(A)$ is the {\it closure} of  $A$ (with respect to $L$).

Then the preceding Note implies the following.

\bc\label{pretypes2types}
Let $A$, $B\in {\bf Types}_T$. Any  cowordism $\sigma$ which is a morphism of types
 $$\sigma: A\to B$$
 is also a morphism of closed types
 $$\sigma: Cl(A)\to Cl(B)B.$$
\ec
{\bf Proof} By the preceding Note, we have a {\it covariant} functor
$$(.)^{\bot\bot}:{\bf Types}_T\to{\bf CTypes}_L.$$
But, under identification (\ref{bidal identification}), it sends any type to its closure and is identity on morphisms. $\Box$
\smallskip

For closed types $A$, $B$ we define the {\it closed tensor product type} $A\otimes B$ as the closure of the tensor product type,
$$A\otimes B=Cl(\{\sigma\otimes \tau|\mbox{ }\sigma\in A,\tau\in B\}).$$

\nb
With the above defined tensor product and duality $(.)^\bot$, the category ${\bf CTypes}_L$ is $*$-autonomous.

 The forgetful functor
 $${\bf CTypes}_L\to {\bf Cow}_T,$$
  which sends type $A$ to the boundary $\d A$ and is identity on morphisms, preserves $*$-autonomous structure. $\Box$
\nbe
\smallskip

It is useful to observe that the original language $L$ can be represented as a closed type of $L$.

Indeed, let $X$ be some boundary with $|X_l|=|X_r|=1$.

Any regular cowordism from ${\bf 1}$ to $X$, seen as a graph consists of a single edge. Define ${\bf star} $ as the type on $X$ consisting of the single regular cowordism whose only edge is labeled with $\star$.

Then the closed type $S={\bf star}^\bot$ consists of all regular cowordisms whose only edge is labeled with an element of $L$. It seems natural to identify $S$ with the language $L$.

\section{Linear logic grammars}
\subsection{Linear logic}
   Strictly speaking, the system discussed below is {\it multiplicative linear logic}, a fragment of full linear logic. However, since we do not consider other fragments, the prefix ``multiplicative'' will be omitted. A more detailed introduction to linear logic can be found in \cite{Girard}, \cite{Girard2}.

Given a set $N$ of  {\it positive literals}, we define the set $N^\bot$ of {\it negative literals}
as
$$N^\bot=\{X^\bot|\mbox{ }X\in N\}.$$
Elements of $N\cup N^\bot$ will be called {\it literals}.

The set  $Fm(N)$ of  ${\bf LL}$ formulas (over the alphabet $N$) is defined by the following induction.
\begin{itemize}
  \item Any $X\in N\cup N^\bot$ is a formula;
  \item if $X$, $Y$ are  formulas, then $X\wp Y$ and $X\otimes Y$ are   formulas;
  \end{itemize}

Connectives $\otimes$ and $\wp$ are called respectively {\it times} (also {\it tensor}) and {\it par} (also {\it cotensor}).

{\it Linear negation} $A^\bot$ of a
formula $A$ is defined
inductively as
$$(P^\bot)^\bot=P,\mbox{ for }P\in N,$$
$$(A\otimes B)^\bot=A^\bot\wp B^\bot,\quad
(A\wp B)^\bot=A^\bot\otimes B^\bot.$$

{\it Linear implication} is defined as
\be\label{linear implication}
A\multimap B=A^\bot\wp B.
\ee

An  ${\bf LL}$ sequent is an expression of the form $\vdash\Gamma$, where $\Gamma$ is s finite sequence of ${\bf LL}$ formulas.

The {\it sequent calculus} for ${\bf LL}$ is given by the following rules:
$$\vdash X^\bot,X\mbox{ } (Identity),\quad
\frac{\vdash \Gamma,X\quad\vdash
X^\bot,\Delta}{\vdash\Gamma,\Delta} \mbox{ }(Cut),$$
$$\frac{\vdash
X_1,\ldots,X_n}{\vdash X_{\pi(1)},\ldots,X_{\pi(n)}},
\pi\in S_n
\mbox{ }(Exchange),$$

$$\frac{\vdash \Gamma,X,Y}
{\vdash\Gamma,X\wp Y}\mbox{ }
(\wp)\quad\frac{\vdash\Gamma, X \quad\vdash
Y,\Delta}{\vdash\Gamma,X\otimes Y,\Delta}\mbox{ }
(\otimes).$$
\smallskip

Linear logic   enjoys the  fundamental property  of {\it cut-elimination}. Any sequent derivable in ${\bf LL}$ is derivable also in the {\it cut-free} system, i.e., without use of the Cut rule.
Moreover, any proof has an essentially unique, up to some permutation of rules,  {\it cut-free form}, which can be found algorithmically.

This allows computational and categorical interpretations  in the {\it proofs-as-programs} or {\it proofs-as-functions}  paradigm.

\subsection{Semantics}
Categorical interpretation of proof theory is based on the idea that formulas should be understood as objects and proofs, as morphisms in a category, while composition of morphisms corresponds to cut-elimination.

In a two-sided sequent calculus, formulas are interpreted as objects in a monoidal category, and a proof of the sequent
$$X_1,\ldots, X_n\vdash X$$
 is interpreted as a morphism of type
 $$X_1\otimes\ldots\otimes X_n\to X.$$
 This includes the case $n=0$, with the usual convention that the tensor of the empty collection of objects is the monoidal unit ${\bf 1}$.

Then the Cut rule corresponds to  composition. A crucial requirement is that the interpretation should be invariant with respect to cut-elimination; a proof and its cut-free form are interpreted the same.

In the case of linear logic, whose sequents are one-sided, the appropriate setting for categorical interpretation   is   {\it $*$-autonomous categories} \cite{Seely}, \cite{Mellies_categorical_semantics}.

In this setting, a proof of the  sequent
$$
\vdash X_1,\ldots, X_n$$
 is interpreted as a morphism of type
 $${\bf 1}\to X_1\wp\ldots\wp X_n.$$
The Cut rule  corresponds to partial pairing, which can be understood as a symmetrized composition.

 A special case of $*$-autonomous categories are compact categories, and, in particular, categories of cowordisms.

Given a $*$-autonomous category ${\bf C}$ and an alphabet $N$ of positive literals, an interpretation of ${\bf LL}$ in ${\bf C}$ consists in assigning to any positive literal $A$ an object $[A]$ of $\bf C$.
The assignment of objects  extends to all formulas in $Fm(N)$ by the obvious induction
$$[A\otimes B]=[A]\otimes [B],\quad [A^\bot]=[A]^\bot.$$

It is quite customary in literature to omit square brackets and denote a formula and its interpretation by the same expression, and  we will follow this practice when convenient.

Given interpretation of formulas, proofs are interpreted by induction on the rules.

The axiom $\vdash A^\bot,A$ is interpreted as  the name
$$\ulcorner \id_{[ A]}\urcorner:{\bf 1}\to [ A]^\bot\wp [ A]$$
 of the identity.

The Cut rule  corresponds to partial pairing, as stated above.

The Exchange rule  corresponds to a symmetry transformation.

The $(\wp)$ rule does  nothing.

The $(\otimes)$ rule is linear distributivity (\ref{linear distributivity}). In the case  of a compact category, in particular the category of cowordisms, the $(\otimes)$ rule  just tensors two morphisms together (up to associativity of tensor product).

Two proofs  are {\it equivalent}, if they get the same interpretation for any interpretation in any $*$-autonomous category.


When the category ${\bf C}$ is a compact category of cowordisms (over some alphabet), and formulas are interpreted as boundaries, we denote the interpretation of a  formula $A$ as $\d A$ and use the convention
$$(\d A)_l=\d_lA,\quad (\d A)_r=\d_r A.$$


Observe that in this,
  interpretations of proofs  do not depend on  the alphabet  at all. So it would be more honest to say that this is  an interpretation  in the category ${\bf Cob}$ of {\it cobordisms}. The alphabet comes into play if we add new axioms to the logic, which gives us a {\it logic grammar}.

\subsection{Adding lexicon}
An ${\bf LL}$ grammar is an interpretation of ${\bf LL}$ in a category of cowordisms supplied with a set of axioms together with cowordisms representing their ``proofs''. Here is an accurate definition

\bd
{\bf Linear logic grammar (LLG)} $G$ is a tuple $G=(N,T,Lex,S)$, where
\begin{itemize}
  \item $N$ is a finite set of positive literals together with an interpretation $A\mapsto \d A$ of elements of $N$ as boundaries;
  \item $T$ is a finite alphabet;
  \item $Lex$, the {\bf lexicon}, is a finite set of expressions of the form $\sigma:F$, where $F$ is an ${\bf LL}$ formula, and $$\sigma:{\bf 1}\to \d F$$ is a cowordism;
  \item $S\in N$, the {\bf standard type}, is interpreted as a boundary with $|\d_lS|=|\d_rS|=1$.
\end{itemize}
\ed
Elements of the lexicon $Lex$ will be often called {\it axioms}, and elements of $N$  will be called {\it atomic types}.

Now let $A$ be an ${\bf LL}$ formula, and let $\rho:{\bf 1}\to \d A$ be a   cowordism.

We say that $G$ {\it generates the cowordism $\rho$  of type $A$}, if there exists axioms
$$\tau_1: A_1,\ldots,\tau_k: A_n\in Lex$$
for some $k\geq 0$ and a cowordism
%
%
$$\sigma:{\bf 1}\to \d A_1^\bot\otimes\ldots\otimes \d A_n^\bot\otimes \d A$$ arising as the interpretation of some ${\bf LL}$ proof  of the sequent
$$\vdash A_1^\bot,\ldots,A_n^\bot, A,$$
such that,
$$\rho=\langle\tau_1\otimes\ldots\otimes\tau_n,\sigma\rangle_{\d A_1\otimes\ldots\otimes \d A_n}.$$

The {\it cowordism type} $A$ {\it generated by $G$}, or, simply, the {\it cowordism type $A$ of $G$}, is the set of all cowordisms of type $A$ generated by $G$.

 Now any regular cowordism of the standard type $S$ is an edge-labeled graph containing  a single edge. Thus the set of type $S$ regular cowordisms can be identified with a set of words.

The {\it language $L(G)$ generated by $G$} is the set of type $S$ regular cowordisms generated by $G$.

\section{Encoding multiple context-free grammars}
In this section,  as an example, we establish a relationship between LLG and {\it multiple context-free grammars}.

\subsection{Multiple context-free grammars}
 Multiple context-free grammars were introduced in \cite{Seki}. We follow (with minor variations in notation) the presentation in \cite{Kanazawa}.

\bd\label{MCFG def} A {\bf multiple context free grammar (MCFG)} $G$ is a tuple $G=(N,T,S,P)$ where
\begin{itemize}
  \item $N$ is a finite  alphabet of nonzero arity predicate symbols  called {\bf nonterminal symbols} or {\bf nonterminals};
  \item $T$ is a finite alphabet of {\bf terminal symbols} or {\bf terminals};
  \item $S\in N$, the {\bf start symbol},  is unary;
  \item $P$ is a finite set of sequents, called {\bf productions}  of the form
  \be\label{production}
   B_1(x^1_1,\ldots,x^1_{k_1}),\ldots,B_n(x^n_1,\ldots,x^n_{k_n})\vdash A(s_1,\ldots,s_k),
  \ee
  where
  \begin{enumerate}[(i)]
    \item $n\geq0$ and $A,B_1,\ldots, B_n$ are nonterminals  with arities $k,k_1,\ldots,k_n$ respectively;
  \item $\{x^j_i\}$ are pairwise distinct variables not from $T$;
  \item $s_1,\ldots, s_k$ are  words built of terminals and $\{x_i^j\}$;
  \item each of the variables $x^j_i$ occurs exactly once in exactly one of the words $s_1,\ldots s_k$.
  \end{enumerate}
 \end{itemize}
\ed
\smallskip

{\bf Remark} Productions are often written in the opposite order in literature; with $A$ on the left and $B_1,\ldots,B_n$ on the right.

Also, our ``non-erasing'' condition (iv) in the definition of a MCFG, namely, that all  $x_i^j$ occurring on the left occur exactly once on the right, is too strong compared with  original definitions in \cite{Seki}, \cite{Kanazawa}. Usually it is required only that each $x_i^j$ should occur at most once on the right. However, it is known \cite{Seki} that adding the non-erasing condition does not change the expressive power of MCFG, in the sense that the class of generated languages (see below) remains the same.

\bd The set of  {\bf predicate formulas derivable in $G$} is the smallest set satisfying  the following. .
\begin{enumerate}[(i)]
 \item {
 If a production $$\vdash A(s_1,\ldots,s_k)$$ is in $P$, then  $A(s_1,\ldots,s_k)$ is derivable.
 }
 \item
 {
 For every production {\bf(\ref{production})} in $P$,
 if
 \begin{itemize}
   \item  $B_1(s^1_1,\ldots,s^1_{k_1}),\ldots,B_n(s^n_1,\ldots,s^n_{k_n})$
are derivable,
   \item $t_m$ is  the result of substituting the word $s^j_i$ for every variable $x^j_i$ in $s_m$,  for  $m=1,\ldots,k$,
 \end{itemize}
 then the formula $A(t_1,\ldots,t_k)$ is derivable.
 }
 %
\end{enumerate}
\ed
\bd
The {\bf language generated } by an MCFG $G$ is the set of words $s$ for which $S(s)$ is derivable in $G$.

{\bf Multiple context-free language} is a language generated by some MCFG.
\ed

When all predicate symbols in $N$ are unary, the above definition reduces to the more familiar case of a {\it context free grammar} (CFG).

\subsection{MCFG productions as cowordisms}\label{MCFG -> cowordisms}
Assume that we are given alphabets $N$ and $T$  of nonterminals and terminals respectively, as in Definition
\ref{MCFG def}.

For each $A\in N$ with arity $k$ introduce  {\it left vertices}
$$l^A_{1},\ldots,l^A_{k}$$ and  {\it right vertices} $$r^A_{1},\ldots,r^A_{k}.$$

Denote the set of left vertices as $\d_lA$, and the set of right vertices, as $\d_r A$.

Define the boundary $\d A$ as $\d A=\d_l A\cup \d_r A$.

Now for any production $p$ of the form
$$   B_1(x_1^1,\ldots,x^1_{k_1}),\ldots,B_n(x^n_1,\ldots,x^n_{k_n})\vdash A(s_1,\ldots,s_k)$$
    we construct a cowordism $graph(p)$ over the alphabet $T$ of the type
  $$graph(p):\d B_1\otimes\ldots\otimes \d B_n\to \d A,$$ if $n>0$,
  or
  $$graph(p):{\bf 1}\to \d A$$ otherwise.

  In order to get $graph(p)$
  it is sufficient to construct a multiword with the left boundary
$$\d_r A\sqcup (\d_l B_1)\sqcup\ldots \sqcup (\d_l B_n)$$
and  the right boundary
$$\d_l A\sqcup (\d_r B_1)\sqcup\ldots \sqcup (\d_r B_n).$$

  The multiword is constructed as follows.

Let $V$ be the set of all variables $x_i^j$ occurring in $p$.

For each $y=x_i^j\in V$ let
$$h(y)=l^{B_j}_i,\quad t(y)=r^{B_j}_i.$$

Now each word $s_m$, $m=1,\ldots,k$, on the righthand side of $p$ is a concatenation of the form
$$s_m=w^0_my^1_mw^1_m\ldots y^{\alpha_m}_mw^{\alpha_m}_m,$$
 where all $$w^0_m,\ldots,w^{\alpha_m}_m$$
 are words in the alphabet $T$ (possibly empty), and
 $$y^1_m,\ldots,y^{\alpha_m}_m$$
  are variables from $V$. (With the convention that $\alpha_m$ may equal zero, in which case $s_m=w^0_m$.)

We represent $p$ as the following multiword $graph(p)$.

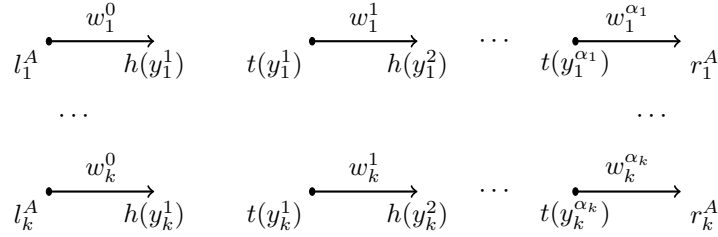
$$
 \tikz[xscale=.7]{

\draw[thick,->](0,0) -- (2,0);
\draw[thick,->](5,0) -- (7,0);
\draw[thick,->](10,0) -- (12,0);
\draw [fill] (0,0) circle [radius=0.05];
\draw [fill] (5,0) circle [radius=0.05];
\draw [fill] (10,0) circle [radius=0.05];
\node at (8.5,0){$\cdots$};
\node[above]at (1,0){$w_1^0$};
\node[above]at (6,0){$w_1^1$};
\node[above]at (11,0){$w_1^{\alpha_1}$};
\node[below left] at (0,0){$l^{A}_1$};
\node[below ] at (2,0){$h(y_1^1)$};
\node[below left]at (5,0){$t(y^1_1)$};
\node[below ]at (7,0){$h(y_1^2)$};
\node[below ]at (10,0){$t(y^{\alpha_1}_1)$};
\node[below right]at (12,0){$r^A_1$};

\node[right] at(0,-1) {$\cdots$};
\node[left] at(12,-1) {$\cdots$};

\draw[thick,->](0,-2) -- (2,-2);
\draw[thick,->](5,-2) -- (7,-2);
\draw[thick,->](10,-2) -- (12,-2);
\draw [fill] (0,-2) circle [radius=0.05];
\draw [fill] (5,-2) circle [radius=0.05];
\draw [fill] (10,-2) circle [radius=0.05];
\node at (8.5,-2){$\cdots$};
\node[above]at (1,-2){$w_k^0$};
\node[above]at (6,-2){$w_k^1$};
\node[above]at (11,-2){$w_k^{\alpha_k}$};
\node[below left] at (0,-2){$l^{A}_k$};
\node[below ] at (2,-2){$h(y_k^1)$};
\node[below left]at (5,-2){$t(y^1_k)$};
\node[below ]at (7,-2){$h(y_k^2)$};
\node[below ]at (10,-2){$t(y^{\alpha_k}_k)$};
\node[below right]at (12,-2){$r^A_k$};
}
$$

In a verbal language, the multiword $graph(p)$ is defined as follows.

For each $m=1,\ldots,k$, if $\alpha_m=0$ draw a directed edge from $l^A_m$ to $r^A_m$ and label it with $s_m$.

Otherwise
\begin{itemize}
  \item draw a directed edge from $l^A_m$ to  $h(y^1_m)$ and label it with $w^0_m$,
  \item draw a directed edge from
$t(y^{\alpha_m}_m)$ to  $r^A_m$ and label it with $w^{\alpha_m}_m$,
  \item for each  $\beta=1,\alpha_m-1$ draw a directed edge from $t(y^\beta_m)$ to $h(y^{\beta+1}_m)$ and label it with $w^\beta_m$.
\end{itemize}

Since each element of $V$ occurs on the left side of $p$ exactly once, it follows that the obtained edge-labeled graph is a perfect matching, hence a (regular) multiword, and its boundary satisfies the desired specification.
\smallskip

 The constructed cowordism  $graph(p)$ represents the production $p$ is a very direct sense.

 Let us construct, for every nonterminal $C\in N$ of arity $\alpha$, an oriented graph on the vertex set $\d C$ by drawing for each $m=1,\ldots,\alpha$ a directed edge from $l^{C}_m$
to $r^{C}_m$ as depicted below.
$$
 
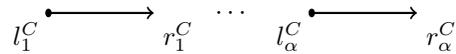
\tikz[xscale=.7]{

\draw[thick,->](0,0) -- (2,0);
\draw[thick,->](5,0) -- (7,0);
\draw [fill] (0,0) circle [radius=0.05];
\draw [fill] (5,0) circle [radius=0.05];
\node at (3.5,0){$\cdots$};
\node[below left] at (0,0){$l^{C}_1$};
\node[below right] at (2,0){$r^{C}_1$};
\node[below left]at (5,0){$l^{C}_\alpha$};
\node[below right]at (7,0){$r^{C}_\alpha$};

}
$$

 This graph is a perfect matching. We call it the {\it pattern } of $C$ and denote as $Pat(C)$.

We will represent a predicate formula
\be\label{just a formula}
C(s_1,\ldots,s_\alpha),
\ee
 where $s_1,\ldots,s_\alpha$ are words, as a multiword whose underlying graph is $Pat(C)$ in the following obvious way.

$$
 \tikz[xscale=.7]{

\draw[thick,->](0,0) -- (2,0);
\draw[thick,->](5,0) -- (7,0);
\draw [fill] (0,0) circle [radius=0.05];
\draw [fill] (5,0) circle [radius=0.05];
\node at (3.5,0){$\cdots$};
\node[above]at (1,0){$s_1$};
\node[above]at (6,0){$s_\alpha$};
\node[below left] at (0,0){$l^{C}_1$};
\node[below right] at (2,0){$r^{C}_1$};
\node[below left]at (5,0){$l^{C}_\alpha$};
\node[below right]at (7,0){$r^{C}_\alpha$};
}
$$

We say that the above multiword  {\it  represents} formula (\ref{just a formula}).

Then the following holds.
 \nb\label{production is a cowordism}
 Let $\sigma_1,\ldots,\sigma_n$ be cowordisms,
 $$\sigma_j:{\bf 1}\to \d B_j,\mbox{ }j=1,\ldots,n,$$
 such that, seen as multiwords, they represent formulas
 $$
 B(s_1^1,\ldots,s_1^{k_1}),\ldots,B(s_n^1,\ldots,s_n^{k_n})
 $$
 respectively, where $k_j$ is the arity of $B_j$, $j=1,\ldots,n$.

 Let $t_m$ be  the result of substituting the word $s^j_i$ for every variable $x^j_i$ in $s_m$,  for  $m=1,\ldots,k$,

 Then the composition $$graph(p)\circ(\sigma_1\otimes\ldots\otimes\sigma_n):{\bf 1}\to \d A$$
 gives the multiword representing the formula
 $$A(t_1,\ldots,t_k).\mbox{ }\Box$$
   \nbe

\subsection{From MCFG to LLG}\label{encoding MCFG}
Any MCFG $G=(N,T,P,S)$ gives rise to an LLG by means of the translation described in Section \ref{MCFG -> cowordisms}.

We treat each nonterminal $A$ as a positive literal and assign to it  the boundary $\d A$ as in Section \ref{MCFG -> cowordisms}. This gives us a set  $ N$ of positive literals and an interpretation $ A\mapsto \d A$ in the category of cowordisms.

Then, to any production $p\in P$ of form (\ref{production}) we assign the  axiom
$$\ulcorner{graph(p)}\urcorner:\d B_1\otimes\ldots\otimes \d B_n\multimap \d A,$$
and this gives us the lexicon $Graph(P)$.

The LLG $G'$ is defined as the tuple $G'=(N,T,Graph( P), S)$.

From Note \ref{production is a cowordism} (using Note \ref{partial pairing} on the properties of partial pairing of cowordisms) it is immediate that the language generated by $G$ identifies with  a subset of the language generated by $ G'$ .

Let us prove the opposite inclusion.

Let $L(G)$ be the language generated by $G$. Consider the category ${\bf CTypes}_{L(G)}$ of closed types of $L(G)$.

For any $A\in N$ of arity $k$ we define the type $ \tilde A$
as the type on $\d A$ consisting of all multiwords representing  formulas
$$A(s_1,\ldots,s_k)$$
derivable in $G$.
We then define the closed type $A\in{\bf Types}_{L(G)}$ as the closure
$$ A=Cl(\tilde A).$$
(We deliberately abuse notation using the same symbol for an atomic type of $G'$ and the corresponding closed cowordism type.)

Now we refine the interpretation of ${\bf LL}$ in ${\bf Cow}_T$ to an interpretation in ${\bf CTypes}_{L(G)}$.

We assign to each literal $A\in  N$ the corresponding cowordism type $A\in{\bf CTypes}_{L(G)}$ and extend the assignment to all formulas in $Fm(N)$ by induction.

Since the category ${\bf CTypes}_{L(G)}$ is $*$-autonomous this gives us  also  a sound interpretation of proofs as morphisms of closed types.

Since the forgetful functor $${\bf CTypes}_{L(G)}\to {\bf Cow}_T$$ preserves $*$-autonomous structure, the two interpretations (in ${\bf CTypes}_{L(G)}$ and in ${\bf Cow}_T$) coincide on the level of cowordisms.
In particular, if $\pi$ is a proof of a sequent $$\vdash A_1,\ldots, A_n,$$
then its interpretation, the cowordism $$[\pi]:{\bf 1}\to \d A_1\otimes\ldots\otimes \d A_n$$
is in the type $A_1\wp\ldots\wp A_n$.

Now we have the following.

\nb
Elements of the type $S\in {\bf CTypes}_{L(G)}$ are all regular cowordisms  whose single edge is labeled with a word of $L$.
\nbe
{\bf Proof} repeats the discussion in the end of Section \ref{cowordism of lang}. $\Box$

\nb
For any axiom $\sigma: F$ in the lexicon  $Graph(P)$, the cowordism $\sigma$ belongs to the corresponding cowordism type $F\in {\bf CTypes}_{L(G)}$.
\nbe
{\bf Proof}
We have that $\sigma=\ulcorner graph(p)\urcorner$ is the name of a cowordism representing some production $p\in P$ of form (\ref{production}), and
$$F=B_1\otimes\ldots\otimes B_n\multimap A.$$

By Note \ref{production is a cowordism}, the cowordism $graph(p)$ is a morphism of types
$$graph(p):\widetilde {B_1}\otimes\ldots\otimes \widetilde B_n\to \widetilde A.$$

By Note \ref{pretypes2types}, it remains a morphism of closed types
$$graph(p): {B_1}\otimes\ldots\otimes  B_n\to  A.$$

It follows that the name $\sigma$ of $graph(p)$ is in the closed type $F$ of $L(G)$. $\Box$
\smallskip

It follows that $G'$  generates the language $L(G)$.
Thus we have the following.

\bt
Any multiple context-free language is generated by an  ${\bf LL}$
 grammar. $\Box$
 \et

\subsection{From LLG to MCFG}
Note that  LLG constructed  from  MCFG in the preceding section have particularly simple lexicons: formulas occurring in such lexicons do not contain $\otimes$ connective. We call such lexicons {\it $\otimes$-free}.

We are going to prove the converse of the preceding theorem: any LLG with a $\otimes$-free lexicon generates a multiple context-free language.

\subsubsection{Extended MCFG grammars}
It will be convenient to reformulate (and slightly generalize) MCFG in a more category-theoretic language.

\bd
An {\bf extended MCFG grammar} $G$ is a tuple $G=(N,T,P,S)$, where
\begin{itemize}
  \item $N$ is a finite set of  {\bf types} together with an interpretation $A\mapsto \d A$ of elements of $N$ as boundaries;
  \item $T$ is a finite alphabet of {\bf terminal symbols};
  \item $P$, is a finite set of rules of the form
\be\label{cowordism produxtion def}
\sigma:\d A_1\otimes\ldots\otimes \d A_n\to \d A,
\ee
Where $A_1,\ldots,A_n,A$ are elements of $N$, and
$$\sigma:\d A_1\otimes\ldots\otimes \d A_n\to \d A.$$
is a cowordism;
  \item $S\in N$, the {\bf standard type}, is interpreted a boundary with $|\d_lS|=|\d_rS|=1$.
\end{itemize}
\ed

Elements of $P$ are called {\it cowordism productions}.

Now, for any type $A\in N$, we will define  a cowordism type on $\d A$, called the {\it cowordism type $A$ generated by $G$}, or, simply, the {\it cowordism type $A$ of} $G$. We will write $G\vdash\sigma:A$ to express that $\sigma $ is in the cowordism type $A$ of $G$.

The set is defined by induction.
\begin{itemize}
  \item If  a cowordism production $\sigma:{\bf 1}\to A$ is in $P$, then $G\vdash\sigma:A$.
  \item If a cowordism production $$\sigma:A_1\otimes\ldots\otimes A_n\to A$$ is in $P$, and $$G\vdash\tau_i:A_i,\quad i=1,\ldots,n,$$
      then $G\vdash\sigma\circ(\tau_1\otimes\ldots\otimes\tau_n):A$.
\end{itemize}

The set of regular cowordisms of type $S$ is called the {\it  language generated by the extended MCFG} $G$.

\subsubsection{From extended MCFG to ordinary MCFG}
Let $G=(N,T,P,S)$ be an  extended MCFG.

For each $A\in N$ and regular cowordism $\sigma:{\bf 1}\to \d A$ such that  $G\vdash\sigma:A$
let $Pat(\sigma)$ be the pattern of $\sigma$.

We say that $Pat(\sigma)$ is a {\it possible pattern} of $A$.

We denote the set of  possible patterns of $A$ as $Patt(A)$. Note that this set is finite.

\bd
The  extended MCFG $G$ is {\bf simple}, if for any  type $A\in N$ the set $Patt(A)$ contains at most one element.
\ed
\smallskip

Quite obviously, any ordinary MCFG, can be seen as a simple extended MCFG.
\bl
If a language is generated by a simple extended MCFG, then it is also generated by an ordinary MCFG.
\el
{\bf Proof} Let $P_0\subseteq P$ be the set of regular cowordism productions that participate in generation of $L(G)$.

For each element $p\in P_0$ we easily write an MCFG production as the inverse of the ``$graph$ map'' (see Section \ref{MCFG -> cowordisms}). This is left as an exercise to the reader. $\Box$
\smallskip

Now we generalise the above to arbitrary extended MCFG $G$.

Since the empty language is obviously multiple context-free, we may assume that $L(G)$ is nonempty, otherwise there is nothing to prove.

We construct a new extended MCFG $G'$ as follows.

For any  type $A$  of $G$ and any   possible pattern $\pi$ of $A$
we introduce a new symbol $(A,\pi)$.

We define the set $N'$ of types of $G'$ as
$$N'=\{(A,\pi)|\mbox{ }A\in N,\pi\in Patt(X)\}.$$
Interpretation of types as boundaries is given by
$$\d (A,\pi)=\d A.$$

For any cowordism production $$\sigma:A_1\otimes\ldots\otimes A_n\to A$$ of $G$ we consider all possible cowordism productions  of the form
\be\label{new lexicon}
\sigma':(A_1,\pi_1)\otimes\ldots\otimes(A_n,\pi_n)\to(A,\pi),
\ee
where
$$\pi_i\in Patt(A_i),\mbox{ } i=1,\ldots, n,$$
 and $\pi\in Patt(A)$ is constructed as the composition
$$\tau=Pat(\sigma)\circ(\pi_1\otimes\ldots\otimes\pi_n).$$

The  set $P'$ of productions for $G'$ consists of all cowordism productions of form (\ref{new lexicon}). Again, there are only finitely many of them.

Since the set $L(G)$ is assumed nonempty,  the set $Patt(S)$ is a singleton. We denote $S'=(S,e)$, where $e$ is the only element of $Patt(S)$.

We define $G'$ as $G'=(N',T,L',S')$.

It is immediate that $G'$ is simple and generates the same extended language as $G$.

Combining the above with the
preceding lemma, we obtain the following.
\bl\label{MCFG=cowordisms}
A language is generated by an MCFG iff it is generated by an  extended MCFG. $\Box$
\el

\subsubsection{From  $\otimes$-free lexicon to extended MCFG }
We start with some simple technical developments.

For a sequent $\Theta$ of the form
\be\label{tensor of ids}
\Theta= A,A^\bot\otimes B^\bot,B,
\ee
we have a proof
$$\frac{\vdash A,A^\bot\quad\vdash B^\bot,B}{\vdash\Theta}(\otimes).$$
We call this proof the {\it standard proof} of $\Theta$.

Now let $\Phi$ be a finite set of $\otimes$-free ${\bf LL}$ formulas, which is closed under subformulas. Let $\Phi^\bot$ be the set
$$\Phi^\bot=\{F^\bot|\mbox{ }F\in\Phi\}.$$

Let $\Pi_0(\Phi)$ be the set of all standard proofs of sequents of form (\ref{tensor of ids}) where $A^\bot,B^\bot,A\wp B\in \Phi$. Let $\Pi(\Phi)$ be the closure of $\Pi_0(\Phi)$  under the Exchange rule.

\bl\label{main}
Let $\Gamma$ be a sequent all whose formulas are in $\Phi^\bot$.

Then any proof of $\Gamma$ is equivalent to a proof obtained from elements of $\Pi(\Phi)$ using only axioms and the Cut rule.
\el
 {\bf Proof} by induction on a cut-free proof. $\Box$
\smallskip

Now let $G=(N,T,Lex,S)$ be an LLG with a $\otimes$-free lexicon.

We construct a cowordism grammar $G'$ using Lemma \ref{main}
as follows.

Let $\Phi$ be the set of all subformulas occurring in $L$.

For every formula $F$ in $\Phi\cup\Phi^\bot$ we introduce a fresh  symbol $[F]$ and assign to $[F]$  the same interpretation  as to $F$,
$$\d[F]=\d F.$$

We put
$$N'=\{[F]\mbox{ }|\mbox{ }F\in\Phi\cup\Phi^\bot\},\quad
S'=[S].$$

Now in order to define an extended MCFG we only need productions.

Let  $P_0$ be the set  of all cowordism productions of the form
$$\sigma:[F_1]\otimes[F_2]\to[F],$$
where $\sigma$ is the interpretation of some proof in $\Pi(\Phi)$ having the sequent
$$\vdash F_1^\bot,F_2^\bot,F$$
as the conclusion.

Let  $P_1$ be the set of all cowordism productions
$$\sigma:{\bf 1}\to[F]$$
where $\sigma:F\in Lex$.

We define the set of productions $P'$ as $P'=P_0\cup P_1'$.

The extended MCFG $G'$ is defined as $G'=(N',T,P',S')$.

Lemma \ref{main} easily yields the following.
\nb
For any formula $F\in \Phi^\bot$ the cowordism type $[F]$ generated by $G'$ coincides with the cowordism type $F$ generated by $G$.
\nbe
{\bf Proof} Exercise. $\Box$
\smallskip

We leave it as an exercise to the reader to prove that if $G$ generates a nonempty language then $S^\bot$ occurs as a subformula in $Lex$, hence $S\in \Phi^\bot$.

Then the above Note implies that the  language of $G'$ coincides with the  language of $G$.

We summarize in the following.
\bl\label{LL->cowordisms lemma}
For any LLG $G$ with a $\otimes$-free lexicon there exists a cowordism grammar $G'$ generating the same extended
 language. $\Box$
\el
\smallskip

Putting Lemmas \ref{LL->cowordisms lemma} and  \ref{MCFG=cowordisms}  together we obtain the following.
\bt\label{tens. free lex is MCFG}
A language is multiple context-free iff it is generated by an LLG with a $\otimes$-free lexicon.
$\Box$
\et

\section{Encoding abstract categorial grammars}
Abstract categorial grammars (ACG) were introduced in \cite{deGroote}. They are based on the purely implicational fragment of linear logic, and {\bf LL} grammars of this paper can be seen as a representation and extension of ACG (over string signature).

In this section we assume that the reader is familiar with basic notions of  $\lambda$-calculus, see \cite{Barendregt} for a reference.

\subsection{Linear $\lambda$-calculus}
Linear $\lambda$-terms are $\lambda$-terms where each  variable occurs exactly once.

More accurately,   given a  set $X$ of {\it variables} and a   set $C$ of  {\it constants}, with $C\cap X=\emptyset$, the set $\Lambda(X,C)$ of {\it linear $\lambda$-terms} is defined by the following.
\begin{itemize}
\item Any $a\in X\cup C$ is in $\Lambda(X,C)$;
\item if $t,s\in\Lambda(X,C)$ are linear $\lambda$-terms whose sets of free variables are disjoint then $(ts)\in\Lambda(X,C)$;
\item if $t\in\Lambda(X,C)$, and $x\in X$ occurs freely in $t$ exactly once then $(\lambda x.t)\in \Lambda(X,C)$.
\end{itemize}

We type linear terms using {\it linear implicational types}.

Given a  set $N$ of {\it atomic types}, the set $Tp(N)$ of { linear implicational types} is defined by induction.
\begin{itemize}
\item Any $A\in N$ is in $Tp(N)$;
\item if $A,B\in Tp(N)$, then $(A\multimap B)\in Tp(N)$.
\end{itemize}

\bd
A {\bf higher order linear signature}, or, simply, a {\bf signature}, $\Sigma$ is a triple $\Sigma=(N,C,\tau)$, where $N$ is a finite set of atomic types, $C$ is a finite set of constants and $\tau$ is a function assigning to each constant a linear implicational type.
\ed

Given a signature $\Sigma=(N,C,\tau)$ and a countable set $X$ of variables, a {\it typing judgement} is a sequent of the form
$$x_1:A_1,\ldots,x_n:A_n\vdash_\Sigma t:A,$$
where $x_1,\ldots x_n\in X$ are pairwise distinct ($n$ may be zero), $t\in\Lambda(X,C)$, and $A_1,\ldots,A_n,A\in Tp(N)$.

 Typing judgements are derived from the following type inference rules.
$$\frac{}{\vdash_\Sigma c:\tau(c)},\mbox{ for }c\in C\quad(\mbox{const}), \quad\frac{}{x:A\vdash_\Sigma x:A}\quad(\mbox{var}),$$
$$\frac{\Gamma\vdash_\Sigma s:A\quad\Delta\vdash_\Sigma t:A\multimap B}{\Gamma,\Delta\vdash_\Sigma (ts):B}\quad(\mbox{app}), \quad\frac{\Gamma,x:A,\Delta\vdash_\Sigma t:B}{\Gamma,\Delta\vdash_\Sigma (\lambda x.t):A\multimap B}\quad(\mbox{abstr}).$$

We say that a term $t$ is {\it typeable} in $\Sigma$ if there is a type $A$ such that $\vdash_\Sigma t:A$. In this case we say that $A$ is the {\it type of $t$ in $\Sigma$}.

\subsubsection{Semantics}
Let ${\bf C}$ be a symmetric monoidal category, and $\Sigma=(N,C,\tau)$ a signature.

An interpretation of signature $\Sigma$ types  in ${\bf C}$ consists in assigning to each atomic type $A\in N$ an object $[A]\in{\bf C}$. This is extended to all types in $Tp(N)$ by the obvious induction:
$$[A\multimap B]=[A]\multimap [B].$$

In the following we omit square brackets and denote a type $A\in Tp(N)$ and its interpretation the same.

An interpretation of $\Sigma$ in ${\bf C}$ consists of an interpretation of types and a function $c\mapsto[c]$ assigning to each constant $c\in C$ a morphism
$$[c]:{\bf 1}\to \tau(c).$$

The interpretation extends to all typeable terms and derivable typing judgements.

To each derivable typing judgement $\sigma$ of the form
$$x_1:A_1,\ldots,x_n;A_m\vdash_\Sigma A$$
we assign a ${\bf C}$-morphism
$$[\sigma]:
A_1\otimes\ldots\otimes A_n\to A,$$
if $n>0$, or
$$[\sigma]:{\bf 1}\to A,$$ if $n=0$, by induction on type inference rules.

If the judgement $\sigma$ is $\vdash_\Sigma c:\tau(c)$  obtained by the (const) rule, then $[\sigma]=[c]$.

If $\sigma$ is $x:A\vdash_\Sigma x:A$  obtained by the (var) rule, then $[\sigma]=\id_{A}$.

If $\sigma$ is obtained from a derivable judgement $\sigma'$ by the (abstr) rule, then $[\sigma]$ is obtained from $[\sigma']$ using symmetry and  correspondence (\ref{monoidal closure}).

If $\sigma$ is obtained from  derivable judgements
$$\sigma_1=\Gamma_1\vdash_\Sigma s:A,\quad \sigma_2=\Gamma_2\vdash_\Sigma t:A\multimap B$$
 by the (app) rule, then
$$[\sigma]=\ev_{A,B}\circ([\sigma_1]\otimes[\sigma_2]).$$

Finally, for a typeable term $t$ of type $A$ we have a derivable typing judgement $\vdash_\Sigma t:A$, and we put
$[t]=[\sigma]$.

\bl\cite{HylandDePaiva}
With notation as above we have:
\begin{itemize}
\item if typeable terms $t,s$ are $\beta\eta$-equivalent, then $[t]=[s]$;
\item if $\vdash_\Sigma s:A$, $\vdash_\Sigma t:A\multimap B$, then $[ts]=[t]\cdot[s]$.
\end{itemize}
\el
{\bf Proof} Exercise or see \cite{HylandDePaiva}. $\Box$

\subsubsection{String signature}
Let $T$ be a finite alphabet.

The {\it string signature} $Str_T$ {\it over} $T$ has a single atomic type $O$, the alphabet $T$ as the set of constants and a  typing assignment
$$\tau(c)=O\multimap O\mbox{ }\forall c\in T.$$

We denote the type $O\multimap O$ as $str$.

Terms typeable in $Str_T$ with the type $str$ are called
{\it string terms}.

Any word $a_1\ldots a_n$ in the alphabet $T$ can be represented as the string term
$$/a_1\ldots a_n/=(\lambda x.a_1(\ldots(a_n(x))\ldots)).$$

It is not hard to see that, if we identify $\beta\eta$-equivalent terms, the map $w\mapsto/w/$ has an inverse.
\nb
Any $\beta$-normal term $t$ typeable in $Str_T$ with the type $str$ is $\beta\eta$-equivalent to the term $/w/$ for some $w\in T^*$.
\nbe
{\bf Proof}
\begin{enumerate}[(i)]
\item There is no typeable term of type $O$ (for example, because any derivable typing judgement has an even number of $O$ occurrences).
\item Using (i), we prove by induction on type inference that any $\beta$-normal term $t$ typeable in $Str_T$ is either a constant $t\in T$, or an abstraction, $t=(\lambda x.t')$ for some variable $x$ and term $t'$.
\item Using (ii), we prove by induction on type inference that for any derivable typing judgement $x:O\vdash_{Str_T}t:O$, where $t$ is a $\beta$-normal term, it holds that $t=c_1(\ldots(c_n(t))\ldots)$ for some constants $c_1,\ldots,c_n\in T$.
\end{enumerate}

Now if $\vdash_{Str_T}t:O\multimap O $, then either $t$ is a constant, hence  $\beta\eta$-equivalent to $/t/$, or its typing was obtained by the (abstr) rule. In the latter case the claim follows from (iii). $\Box$
\smallskip

Thus we have a map from typeable string terms to words over $T$.
It turns out that this map extends to all typeable terms as a map to cowordisms.

Let us choose an interpretation of the atomic type $O$ as a one-point boundary $$\d O=\d_lO\cup\d_rO$$ with $|\d_rO|=1$, $\d_lO=\emptyset$.

By induction this gives us an interpretation $A\mapsto\d A$ of all types in $Tp(O)$ as boundaries.

We extend this to an interpretation of the string signature in the category ${\bf Types}_T$ by defining the cowordism type $O$ on the boundary
$\d O$ as the empty set.

Any regular cowordism $\sigma:\d O\to \d O$ which is a  morphism of types $\sigma:O\to O$, is a graph consisting of a single edge labeled with some word $w\in T^*$. We denote this cowordism as $graph(w)$.

We interpret each constant $c\in T$ as the corresponding regular cowordism $graph(c):O\to O$.

This gives us an interpretation of the signature $Str_T$.

 We denote the interpretation of a typeable term $t\in\Lambda(X,C)$
as $graph(t)$.
Note that for any word $w\in T^*$ we have $graph(/w/)=graph(w)$.

We call an interpretation of the above form a {\it standard interpretation of the string signature}.

\subsection{Abstract categorial grammars}
Given two signatures $\Sigma_i=(N_i,C_i,\tau_i)$, $i=1,2$, a {\it map of signatures} $$\phi:\Sigma_1\to\Sigma_2$$ is a
pair $\phi=(F,G)$, where
\begin{itemize}
  \item $F:Tp(\Sigma_1)\to  Tp(\Sigma_2)$ is a function satisfying the homomorphism property $$F(A\multimap B)=F(A)\multimap F(B),$$
  \item $G:C_1\to\Lambda(X,C_2)$ is a function such that
       for any $c\in C_1$ it holds that $\vdash_{\Sigma_2}G(c):F(\tau(c))$.
\end{itemize}

The map $G$ above extends inductively to a map
$$G:\Lambda(X,C_1)\to\Lambda(X,C_1)$$
  by
  $$G(x)=x,\mbox{ }x\in X,$$
   $$G(ts)=(G(t)G(s)),\quad G(\lambda x.t)=(\lambda x.G(t)).$$

For economy of notation, we write $\phi(A)$ for $F(A)$ when $A\in Tp(C_1)$, and we write $\phi(t)$ for $G(t)$ when $t\in\Lambda(X,C_1)$.

\bd
A {\bf string abstract categorial grammar (string ACG)} $G$ is a tuple $G=(\Sigma,T, \phi,S)$, where
\begin{itemize}
  \item $\Sigma$, is a signature;
  \item $T$ is a finite alphabet
  \item $\phi:\Sigma\to Str_T$, the {\bf lexicon}, is a map of signatures;
  \item $S$, {\bf the standard type}, is an atomic type of $\Sigma$, such that $\phi(S)=str$.
\end{itemize}
\ed

The {\it string language $L(G) $ generated by $G$} is the set of words over $T$ given by
$$
L(G)=\{w\in T^*|\mbox{ }\exists t\mbox{ }\vdash_\Sigma t:S\mbox{ and }\phi(t)=/w/\}.
$$
Equivalently
$$
L(G)=\{w\in T^*|\mbox{ }\exists t\mbox{ }\vdash_\Sigma t:S\mbox{ and }graph(t)=graph(w)\}.
$$

\subsection{Encoding}
Let $G=(\Sigma,T, \phi,S)$ be a string ACG.

Choose some standard interpretation of $Str_T$ in ${\bf Types}_T$. This yields us  an interpretation of the signature $\Sigma$ defined as follows.

To any type $A\in Tp(\Sigma)$ we assign the boundary
$$\d A=\d(\phi(A))$$
and the cowordism type $A\in {\bf Types}_T$ given by
$$A=\{graph(\phi(t))|\mbox{ }\vdash_\Sigma t:A\}.$$

To any term $t$ typeable in $\Sigma$ we assign the cowordism $$graph(t)=graph(\phi(t)).$$

It is immediate from definitions that the interpretation is sound, i.e. we have the following.
\nb\label{soundeness of ACG encoding} If $\vdash_\Sigma t:A$ then $graph(t)\in A$. $\Box$
\nbe
\smallskip

Now treating the set of  atomic types of $\Sigma$ as literals and types of $\Sigma$ as ${\bf LL}$ formulas we construct an LLG $G'$ encoding $G$.

Let $N$, $C$ be the sets of, respectively,  atomic types and  constants of $\Sigma$. We already have the assignment $A\mapsto\d A$ of elements of $N$ to boundaries.

We define the set of axioms
$$Lex=\{graph(c):\tau(c)|\mbox{ }c\in C\}.$$

The LLG $G'$ is defined as $G'=(N,T,Lex,S)$.

Now, by induction on type inference rules using Note \ref{soundeness of ACG encoding} we prove that the language $L(G)$ generated by $G$ is a subset of language of $G'$.

Proof of the opposite inclusion  repeats the argument in Section \ref{encoding MCFG} where we consider encoding of MCFG.  We consider the category ${\bf CTypes}_{L(G)}$ of closed types of $L(G)$ and observe that any cowordism type $A$ of $G'$ is a subset of the corresponding closed type  of ${\bf CTypes}_{L(G)}$.

We summarise.
\bt\label{encoding ACG}
If a language is generated by a string ACG then it is also generated by an LLG. $\Box$
\et
\smallskip

It seems an interesting question whether the converse is true or not.
\smallskip

{\bf Remark} Since MCFG embed into string ACG \cite{deGrootePogodalla}, Theorem \ref{encoding ACG} on encoding  ACG in LLG grammars implies that MCFG embed into LLG. However it does not imply  the converse statement (Theorem \ref{tens. free lex is MCFG}, that any $\otimes$-free lexicon gives rise to an MCFG).

 On the other hand it is not hard to see that Theorem \ref{tens. free lex is MCFG} together with Theorem \ref{encoding ACG} do imply the known result \cite{Salvati} that any second order string ACG generates a multiple context-free language. Thus we gave another, more ``category-theoretic''  proof of this result.

\section{Encoding backpack problem}
It is known that ACG, in general, can generate NP-complete languages. In view of Theorem \ref{encoding ACG} it is no wonder that LLG can generate NP-complete languages as well. In this last section we show how an LLG can generate solutions of the backpack problem.
Our purpose here is mainly  illustrative. We try to convince the reader that the geometric language of cowordisms is indeed intuitive and convenient for analysing language generation.

We consider backpack problem in the form of the {\it subset sum problem}.

\bd {\bf Subset sum problem (SSP)}: Given a finite sequence $s$ of integers, determine if there is a subsequence $s'\subseteq s$ such that $\sum\limits_{z\in s}z=0$.
\ed

SSP is known to be NP-complete, see \cite{Martello}.

We now define a language representing solutions of SSP.

We  represent integers as words in the alphabet $\{+,-\}$, we call them {\it numerals}. An integer $z$ is represented (non-uniquely) as a word for which the difference of $+$ and $-$ occurrences equals $z$.

We say that a numeral is {\it irreducible}, if it consists only of pluses or only of minuses.

We represent finite sequences of integers as words in the alphabet $T=\{+,-,\bullet\}$, with $\bullet$ interpreted as a separation sign. Thus a word in this alphabet should be read as a list of numerals separated by bullets.

When all numerals in the list are irreducible, we say that the list is irreducible. Note that any sequence of integers has unique representation as an irreducible list.


We now construct a system of cowordisms  over $T$ which (together with symmetry transformations) generates solutions of SSP.

We will use four atomic boundaries $E,P,H,S$, each of them having one point in the left boundary and one point in the right boundary.

First we construct a system which generates lists of numerals representing sequences that sum to zero.

We define  four cowordisms
$$cons:S\otimes S\to S,\quad open:H\to S$$
$$ push:H\otimes H\to H\otimes H,\quad close:{\bf 1}\to H$$

in the graphical language as follows.
 $$
 
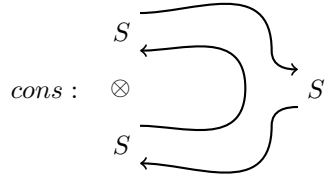
\tikz[xscale=.7]{

 \draw[thick,->](0,0) to  [out=0,in=90] (2.5,-0.5) to  [out=-90,in=180] (3,-0.75);

\draw[thick,<-](0,-0.5) to  [out=0,in=90] (2,-1) to  [out=-90,in=0] (0,-1.5);
  \draw[thick,<-](0,-2.)to  [out=0,in=-90](2.5,-1.5) to  [out=90,in=-180] (3,-1.25);
  \node[above left] at (0,-0.5) {$S$};
\node[above left] at (0,-2) {$S$};
\node[above right] at (3,-1.25) {$S$};
\node[above left] at (0,-1.25) {$\otimes$};
\node[above left] at(-1,-1.25) {$cons:$}
 }$$

$$
 \tikz[xscale=.7]{

\draw[thick,->](0,0) -- (1,0);
\draw[thick,<-](0,-0.5) -- (1,-0.5);
\node[above right] at (1,-.5) {$S$};
\node[above left] at (0,-.5) {$H$};
\node[above left] at(-1,-.5) {$open:$};
 }$$

 $$
 \tikz[xscale=.7]{

\draw[thick,->](0,0) -- (1,0);
\draw[thick,<-](0,-0.5) -- (1,-0.5);
\node[above right] at (1,-.5) {$H$};
\node[above left] at (0,-.5) {$H$};
\node[above] at (0.5,0) {$+$};

\draw[thick,->](0,-1.5) -- (1,-1.5);
\draw[thick,<-](0,-2.) -- (1,-2.);
\node[above right] at (1,-2) {$H$};
\node[above left] at (0,-2) {$H$};
\node[above] at (0.5,-1.5) {$-$};

\node[above right] at (1,-1.25) {$\otimes$};
\node[above left] at (0,-1.25) {$\otimes$};
\node[above left] at(-1,-1.25) {$push:$};
 }$$

   $$
 \tikz[xscale=.7]{
 \node[above] at (0.6,0) {$\bullet$};
\draw[thick,<-](1,0) to  [out=-180,in=90] (0,-0.25) to  [out=-90,in=-180] (1,-.5);

\node[above right] at (1,-.5) {$H$};
\node[above left] at(-1,-.5) {$close:$};
 }$$

The cowordism $cons$, by iterated compositions with itself, generates lists with arbitrary many empty slots. Then the cowordism $open$ converts them into slots that can be filled with pluses and minuses. Then $push$ fill the slots (always in pairs), and  $close$ closes them.

It is easy to see that all cowordisms from ${\bf 1}$ to $S$ generated by the above system (together with symmetry transformations) represent sequences of integers summing to zero, and vice versa, for any sequence summing to zero, its irreducible list representation is generated by the above.

Now, in order to generate solutions of SSP we need some extra ``deceptive'' slots, which  contain elements not summing to zero. These slots  will be represented by the boundary $P$.

%

We define  cowordisms
$$open_P:P\otimes S\to S,\quad close_P:{\bf 1}\to P,$$
$$\quad push_+:P\to P,\quad push_-:P\to P$$
as follows.
%

 $$
 
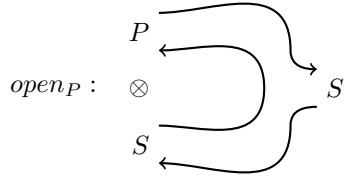
\tikz[xscale=.7]{

 \draw[thick,->](0,0) to  [out=0,in=90] (2.5,-0.5) to  [out=-90,in=180] (3,-0.75);

\draw[thick,<-](0,-0.5) to  [out=0,in=90] (2,-1) to  [out=-90,in=0] (0,-1.5);
  \draw[thick,<-](0,-2.)to  [out=0,in=-90](2.5,-1.5) to  [out=90,in=-180] (3,-1.25);
  \node[above left] at (0,-0.5) {$P$};
\node[above left] at (0,-2) {$S$};
\node[above right] at (3,-1.25) {$S$};
\node[above left] at (0,-1.25) {$\otimes$};
\node[above left] at(-1,-1.25) {$open_P:$}
 }$$

 $$
 \tikz[xscale=.7]{
\draw[thick,->](0,0) -- (1,0);
\draw[thick,<-](0,-0.5) -- (1,-0.5);
\node[above right] at (1,-.5) {$P$};
\node[above left] at (0,-.5) {$P$};
\node[above] at (0.5,0) {$+$};
\node[above left] at(-1,-.5) {$push_+:$}
 }$$

  $$
 \tikz[xscale=.7]{

\draw[thick,->](0,0) -- (1,0);
\draw[thick,<-](0,-0.5) -- (1,-0.5);
\node[above right] at (1,-.5) {$P$};
\node[above left] at (0,-.5) {$P$};
\node[above] at (0.5,0) {$-$};
\node[above left] at(-1,-.5) {$push_-:$};
 }$$

 $$
 \tikz[xscale=.7]{
 \node[above] at (0.6,0) {$\bullet$};
\draw[thick,<-](1,0) to  [out=-180,in=90] (0,-0.25) to  [out=-90,in=-180] (1,-.5);

\node[above right] at (1,-.5) {$P$};
\node[above left] at(-1,-.5) {$close_P:$};
 }$$

%
%
%
%
The cowordism $open_P$ adds deceptive slots to the list, $push_-$ and $push_+$ fill them with arbitrary numerals, and $close_P$ closes them.

%
%
%
%
%
%

Let us denote the set of cowordisms from ${\bf 1}$ to $S$ generated by the above system and symmetry as $L_0$.

It is easy to see that  $L_0$ membership problem is essentially equivalent SSP.
In particular,  a sequence $s$ of integers is a solution of SSP iff the corresponding irreducible list is in $L_0$.
It follows that $L_0$ is NP-hard.

It is also easy to show that $L_0$ membership problem is itself in NP, hence $L_0$ is, in fact, NP-complete.

Finally, observe that if we define an LLG $G$ by a lexicon consisting of names of the above cowordisms, then $G$ will generate $L_0$. This is a technical and not difficult exercise in multiplicative linear logic proof-search.

 \end{document}